\documentclass{acm_proc_article-sp}
\usepackage[us,12hr]{datetime}
\usepackage{mathtools}
\usepackage{algorithm}
\usepackage{algpseudocode}
\usepackage{hyperref}
\usepackage{comment}
\usepackage{color}

\DeclareMathOperator{\sgn}{sgn}
\usepackage{subcaption}

\begin{document}


%
%
%
%

%

\title{Transferability in Machine Learning: from Phenomena to Black-Box Attacks using Adversarial Samples}
%
%
%
%
%

\numberofauthors{2} 
%
\author{
%
\alignauthor
Nicolas Papernot and Patrick McDaniel\\
       \affaddr{The Pennsylvania State University}\\
       \affaddr{University Park, PA}\\
	   \email{\{ngp5056,mcdaniel\}@cse.psu.edu}
\alignauthor
Ian Goodfellow\\
	\affaddr{OpenAI}\\
	\affaddr{San Francisco, CA}\\
	\email{ian@openai.com}
}


\maketitle
\begin{abstract}

Many machine learning models are vulnerable to
{\em adversarial examples}: inputs that are specially crafted to cause
a machine learning model to produce an incorrect output.
Adversarial examples that affect one model often affect another model,
even if the two models have different architectures or were trained
on different training sets, so long as both models were trained to
perform the same task.
An attacker may therefore train their own {\em substitute} model,
craft adversarial examples against the substitute, and {\em transfer}
them to a victim model, with very little information about the victim.
Recent work has further developed a technique that uses the victim model
as an oracle to label a synthetic training set for the substitute, so
the attacker need not even collect a training set to mount the attack.
We extend these recent techniques using {\em reservoir sampling} to greatly
enhance the efficiency of the training procedure for the substitute model.
We introduce new transferability attacks between previously unexplored
(substitute, victim) pairs of machine learning model classes, most notably
SVMs and decision trees.
We demonstrate
our
attacks on two commercial machine learning
classification systems from Amazon (96.19\% misclassification rate) and
Google (88.94\%) using only $800$ queries of the victim model, 
thereby showing that 
existing machine learning approaches
are {\it in general} vulnerable to systematic black-box attacks regardless
of their structure.

\end{abstract}

%
%
%

%
%

%
%




\section{Introduction}

\begin{figure}[h] 
	\centering
	\includegraphics[width=\columnwidth]{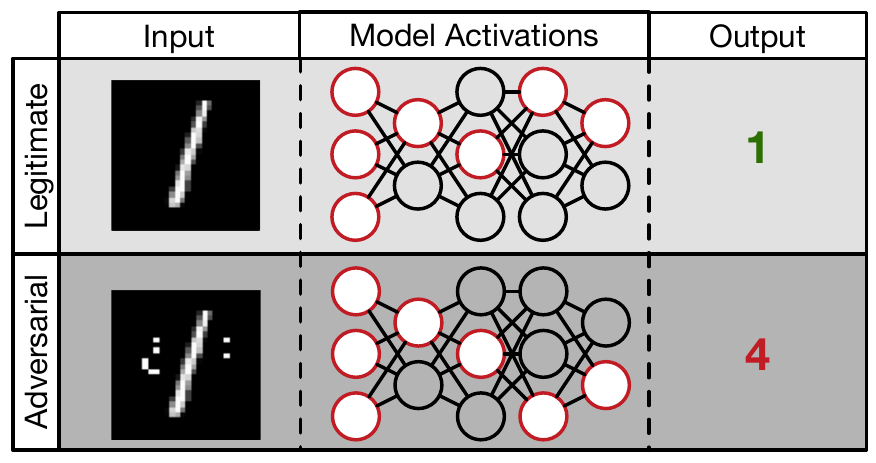} 
	\caption{An adversarial sample (bottom row) is produced by slightly altering a legitimate sample (top row) in a way that forces the model to make a wrong prediction whereas a human would still correctly classify the sample~\cite{papernot2015limitations}.}
	\label{fig:adv-sample-intro}
\end{figure}

Many classes of machine learning algorithms have been shown to be vulnerable to {\it adversarial samples}~\cite{szegedy2013intriguing,goodfellow2014explaining,papernot2015limitations}; adversaries subtly alter legitimate inputs (call input perturbation) to induce the trained model to produce erroneous outputs.  Adversarial samples  can be used to, for example, subvert fraud detection, bypass content filters or malware detection, or to mislead  autonomous navigation systems~\cite{papernot2016practical}.  These attacks on input integrity  exploit imperfections and approximations made by learning algorithms during training to control machine learning models outputs (see Figure~\ref{fig:adv-sample-intro}).  

\emph{Adversarial sample transferability}\footnote{Note that this is distinct from \emph{knowledge transfer}, which refers to techniques designed to transfer the generalization knowledge learned by a model $f$ during training---and encoded in its parameters---to another model $f'$~\cite{hinton2015distilling}.} is the property that some adversarial samples produced to mislead a specific model $f$ can mislead other models $f'$---even if their architectures greatly differ~\cite{szegedy2013intriguing,goodfellow2014explaining,papernot2016practical}.   A practical impact of this property is that it leads to \emph{oracle}-based black box attacks.  In one such attack, Papernot et al.  trained a local deep neural network (DNN)
using crafted inputs and  output labels generated by the target ``victim'' DNN~\cite{papernot2015limitations}.  Thereafter, the local network was used to generate adversarial samples that were highly effective on the original victim DNN.   The key here was that the adversary has very limited information---they knew nothing about the architecture or parameters but only knew that the victim was a DNN---and had only oracle access that allowed it to obtain outputs for chosen inputs.  

In this paper, we develop and validate a generalized algorithm for black box attacks that exploit adversarial sample transferability on broad classes of machine learning.  In investigating these attacks, we explore transferability within and between different classes  of machine learning classifier algorithms. We explore neural networks (DNNs), logistic regression (LR), support vector machines (SVM), decision trees (DT), nearest neighbors (kNN), and ensembles (Ens.).  In this, we demonstrate that black-box attacks are generally applicable to machine learning and can effectively target classifiers not built using deep neural networks.  The generalization is two-fold: we show that (1) the substitute model can be trained with other techniques than deep learning, and (2) transferability-based black box attacks are not restricted to deep learning targets and is in fact successful with targeted models of many machine learning types.  Our contributions are summarized as follows:

\vspace{-1em}
\begin{itemize} 
	
	\item We introduce adversarial sample crafting techniques for support vector machine as well as decision trees---which are non-differentiable machine
	learning models. 
		
	\item We study adversarial sample transferability across the machine learning space and find that samples largely transfer well across models trained with the same machine learning technique, and across models trained with different techniques or ensembles taking collective decisions. For example, a support vector machine and decision tree respectively misclassify $91.43\%$ and $87.42\%$ of adversarial samples crafted for a logistic regression model. 
  Previous work on adversarial example transferability has primarily studied the case where
  at least one of the models involved in the transfer is a neural network \cite{szegedy2013intriguing,goodfellow2014explaining,WardeFarley16},
  while we aim to more generally characterize the transferability between a diverse set of models chosen
  to capture most of the space of popular machine learning algorithms.

   \item We generalize the learning of substitute models from deep learning to logistic regression and support vector machines. Furthermore, we show that it is possible to learn substitutes matching labels produced by many machine learning models (DNN, LR, SVM, kNN) at rates superior to $80\%$. We improve the accuracy and computational cost of a previously proposed substitute learning technique by introducing a new hyper-parameter and the use of reservoir sampling.
	\item We conduct black-box attacks against classifiers hosted by Amazon and Google. We show that despite our lack of knowledge of the
	classifier internals, we can force them to respectively misclassify 96.19\% and	88.94\% of their inputs using a logistic regression substitute model trained by making only $800$ queries to the target. 
	
\end{itemize}


\section{Approach Overview}
\label{sec:transferability-theory}
In this section, we describe our approach, which is structured around the
evaluation of two hypotheses relevant to the design of black-box attacks
against machine learning classifiers. 

Let us precisely define adversarial sample transferability. Consider an
adversary interested in producing an \emph{adversarial sample} $\vec{x^*}$
misclassified in any class different from the class assigned by model $f$ to
legitimate input $\vec{x}$. This can be done by solving\footnote{Finding a
closed form solution to this problem is not always possible, as some machine
learning models $f$ preclude the optimization problem from being linear or
convex. Nevertheless, several approaches have been proposed to find
approximative solutions to
Equation~\ref{eq:adv-sample-crafting-misclassification}. They yield adversarial
samples effectively misleading non-linear and non-convex models like neural
networks~\cite{szegedy2013intriguing,goodfellow2014explaining,papernot2015limitations}.
In addition, we introduce new techniques to craft adversarial samples against
support vector machines and decision trees in
Section~\ref{sec:adv-sample-crafting}.} the following optimization
problem~\cite{szegedy2013intriguing}:
\begin{equation}
\label{eq:adv-sample-crafting-misclassification}
\vec{x^*}=\vec{x}+\delta_{\vec{x}} \texttt{ where } \delta_{\vec{x}} =  \arg\min_{\vec{z}} f(\vec{x}+\vec{z}) \neq f(\vec{x}) 
\end{equation} 
Samples $\vec{x^*}$ solving
Equation~\ref{eq:adv-sample-crafting-misclassification} are specifically
computed to mislead model $f$. However, as stated previously, such adversarial
samples are in practice also frequently misclassified by models $f'$ different
from $f$. To facilitate our discussion, we formalize this \emph{adversarial
sample transferability} notion as:
\begin{equation}
\label{eq:inter-transferability-rate}
\Omega_X(f,f') = \left| \left\{  f'(\vec{x}) \neq f'\left(\vec{x}+\delta_{\vec{x}}\right)  : \vec{x}\in X\right\} \right|
\end{equation}
where set $X$ is representative of the expected input distribution 
for the task solved by models $f$ and $f'$. We partition
 adversarial sample
transferability in two variants  characterizing the pair of
models $(f,f')$. The first, \emph{intra-technique transferability}, is
defined across models trained with the same machine learning technique but
different parameter initializations or datasets (e.g., $f$ and $f'$
are both neural networks or both decision trees). The second,
\emph{cross-technique transferability}, considers models trained using two
 techniques (e.g., $f$ is a neural network and $f'$ a decision tree).

\noindent\textbf{Hypothesis 1:} \emph{Both intra-technique and cross-technique
adversarial sample transferabilities are consistently strong phenomena across
the space of machine learning techniques}. 

In this first hypothesis, we explore how well both variants of transferability
hold across classes of machine learning algorithms. The motivation behind this
investigation is that adversarial sample transferability constitutes a threat
vector against machine learning classifiers in adversarial settings. To
identify the most vulnerable classes of models, we need to generate an accurate
comparison of the attack surface of each class in constrained experimental
settings.  

To validate this hypothesis, we perform a large-scale  study in
Section~\ref{sec:transferability-section}. Each of the
study's two folds investigates one of the adversarial sample transferability variants:
intra-technique and cross-technique. For completeness, we consider a collection of
models representatively spanning the machine learning space, as demonstrated by
Table~\ref{tbl:machine-learning-techniques}. Models are trained on MNIST
data~\cite{lecun1998mnist} to solve the hand-written digit recognition task. In
the first fold of the study, we measure intra-technique adversarial sample
transferability rates $\Omega_X(f,f')$, for each machine learning technique,
across models trained on different subsets of the data.  In the second fold of
the study, we measure inter-technique adversarial sample transferability rates
$\Omega_X(f,f')$ across models corresponding to all possible pairs of machine
learning techniques. 

\begin {table}[t]
\centering
\begin{tabular}{|c||c|c|c|}
	\hline
	ML  & Differentiable & Linear  & Lazy    \\
	Technique & Model & Model & Prediction\\ \hline \hline
	DNN & Yes  & No & No  \\ \hline
	LR  &  Yes & Log-linear & No   \\ \hline
	SVM  & No  & No & No \\ \hline
	DT & No  & No &  No  \\ \hline
	kNN  & No  & No & Yes   \\ \hline
	Ens  & No  & No & No  \\ \hline
\end{tabular}
\caption{Machine Learning Techniques studied in Section~\ref{sec:transferability-section}}
\label{tbl:machine-learning-techniques}
\end{table}

\noindent\textbf{Hypothesis 2:} \emph{Black-box attacks  are possible in
practical settings against any unknown machine learning classifier.}

Our motivation is to demonstrate that deployment of machine learning in
settings where there are incentives for adversaries to have models misbehave
must take into account the practical threat vector of adversarial samples.
Indeed, if black-box attacks are realistic in practical settings, machine
learning algorithm inputs must be validated as being part of the expected
distribution of inputs. As is the case for SQL injections, the existence of
adversarial samples calls for input validation in production systems using
machine learning. 

The verification of this second hypothesis is two-fold as well. In
Section~\ref{sec:learning-approximators}, we show how to transfer the
generalization knowledge  of any machine learning classifiers into a substitute
model by querying the classifier for labels on carefully selected inputs. In
Section~\ref{sec:ml-oracle-attack}, we perform black-box attacks against
commercial machine learning classifiers hosted by Amazon and Google. As we
validate the hypothesis throughout Sections~\ref{sec:learning-approximators}
and~\ref{sec:ml-oracle-attack}, we operate under the specific threat model of
an oracle, described in~\cite{papernot2016practical}, which characterizes
realistic adversarial settings. Instead of having full knowledge of the model's
architecture $f$ and its parameters $\theta$, as was the case for the first
hypothesis validation in Section~\ref{sec:transferability-section}, we now
assume the adversary's only capability is to observe the label predicted by
the model $f$ on inputs of its choice.


\section{Transferability of Adversarial Samples in Machine Learning}
\label{sec:transferability-section}

In this section, our working hypothesis is that intra-technique and cross-technique
adversarial sample transferability are strong phenomena across the machine
learning space. Thus, we empirically study these two phenomena across a range of machine learning techniques: deep
neural networks (DNNs), logistic regression (LR), support vector machines
(SVM), decision trees (DT), nearest neighbors (kNN), and ensembles (Ens.).
All models are found vulnerable to intra-technique adversarial sample
transferability---misclassification of samples by different models trained using
the same machine learning technique, the phenomenon is stronger for
differentiable models like DNNs and LR than for non-differentiable models
like SVMs, DTs and kNNs. Then, we observe that DNNs and kNNs boast resilience
to cross-technique transferability, misclassifications of adversarial samples by
models trained with distinct machine learning techniques. We find that all other models, including LR, SVMs, DTs, and an ensemble of
 models collectively making predictions, are considerably more vulnerable to
cross-technique transferability.

\subsection{Experimental Setup}
\label{sec:transferability-experimental-setup}
We describe here the dataset and machine learning models used in this section
to study both types of transferability.

\textbf{Dataset -} We use the seminal MNIST dataset of handwritten
digits~\cite{lecun1998mnist}. This dataset has been well-studied in both the
machine learning and security communities. We chose it because its
dimensionality is suitable to the range of machine learning techniques included
in our study, which all perform at least reasonably well on this dataset. The
task associated with the dataset is classification of images in one of the $10$
classes corresponding to each possible digit ranging from $0$ to $9$. The
dataset includes $50,000$ training samples, $10,000$ validation samples, and
$10,000$ test samples. Each $28$x$28$ gray-scale pixel image is encoded as a
vector of intensities whose real values range from $0$ (black) to $1$ (white).

\textbf{Machine learning models -} We selected five machine learning
techniques: DNNs, LR, SVMs, DTs, and kNNs. All of these machine learning
techniques, as well as the algorithms used to craft adversarial samples, are
presented in Section~\ref{sec:adv-sample-crafting} of this paper. As outlined
in Table~\ref{tbl:machine-learning-techniques}, DNNs were chosen for their
state-of-the-art performance, LR for its simplicity, SVMs for their potential
robustness stemming from the margin constraints when choosing decision
boundaries at training, DTs for their non-differentiability, and kNNs for being
lazy-classification\footnote{No model is learned during training. Predictions
are made by finding $k$ points closest to the sample in the training data, and
extrapolating its class from the class of these $k$ points.} models. To train
DNN, LR, and kNN models, we use Theano~\cite{bergstra2010theano} and
Lasagne~\cite{lasagne}. The DNN is made up of a hierarchy of 2 convolutional
layers of $32$ $3$x$3$ kernels, 2 convolutional layers of $64$ $3$x$3$ kernels,
2 rectified linear layers of $100$ units, and a softmax layer of $10$ units. It
is trained during 10 epochs with learning, momentum, and dropout rates of
respectively $10^{-2}$, $0.9$, and $0.5$ decayed by $0.5$ after 5 epochs. The
LR is performed using a softmax regression on the inputs. It is trained during
15 epochs at a learning rate of $10^{-2}$ with a momentum rate of $0.9$ both
decayed by $0.5$ after $10$ epochs. The linear SVM and DT are  trained with
scikit-Learn.

\subsection{Intra-technique Transferability}
\label{sec:intra-technique-transferability}

We show that differentiable models like DNNs and LR are more vulnerable to
intra-technique transferability than  non-differentiable models like SVMs, DTs, and
kNNs. We measure \emph{intra-technique transferability} between models $i$ and $j$,
both learned using the same machine learning technique, as the proportion of
adversarial samples produced to be misclassified by model $i$ that are
misclassified by model $j$. 

\begin{figure}[t!] 
	\begin{subfigure}[b]{0.5\columnwidth}
	  \centering
	  \includegraphics[width=\columnwidth]{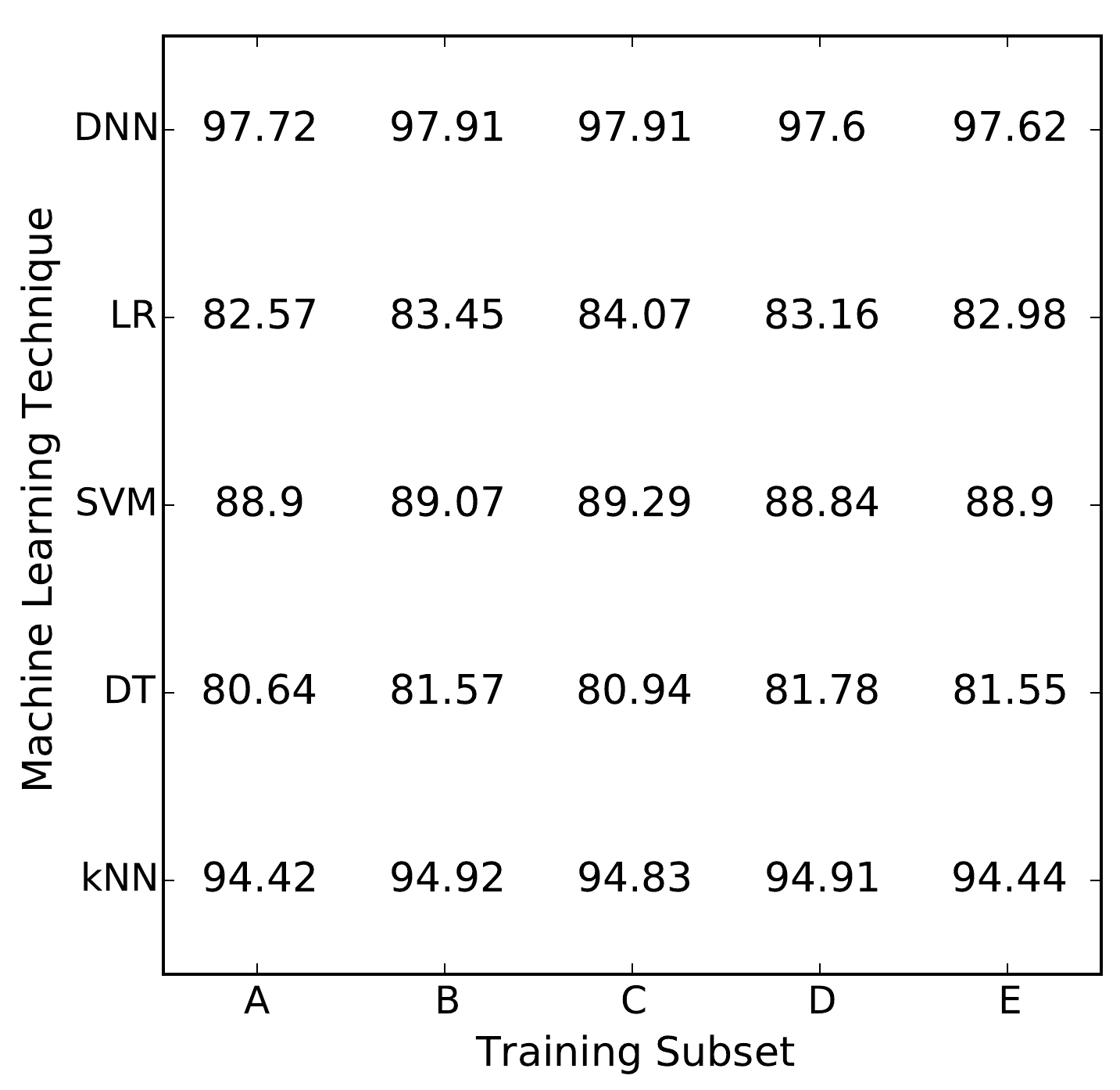} 
	  \caption{Model Accuracies} 
	  \label{tbl:accuracies-intramodel} 
	  \vspace{4ex}
	\end{subfigure}
  \begin{subfigure}[b]{0.5\columnwidth}
    \centering
    \includegraphics[width=\columnwidth]{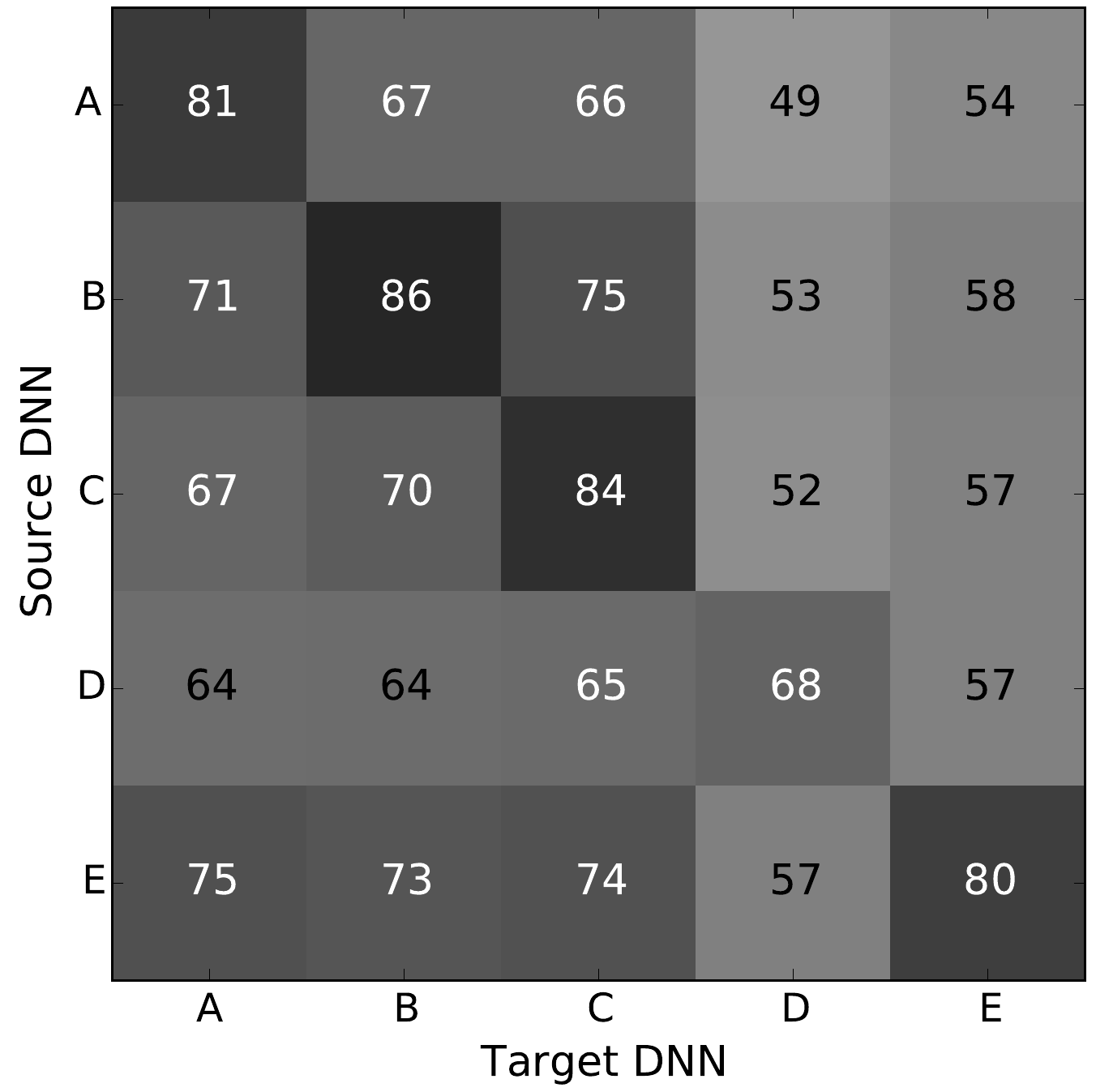} 
    \caption{DNN models} 
    \label{fig:intramodel-transferability:b} 
    \vspace{4ex}
  \end{subfigure}
  \begin{subfigure}[b]{0.5\columnwidth}
    \centering
    \includegraphics[width=\columnwidth]{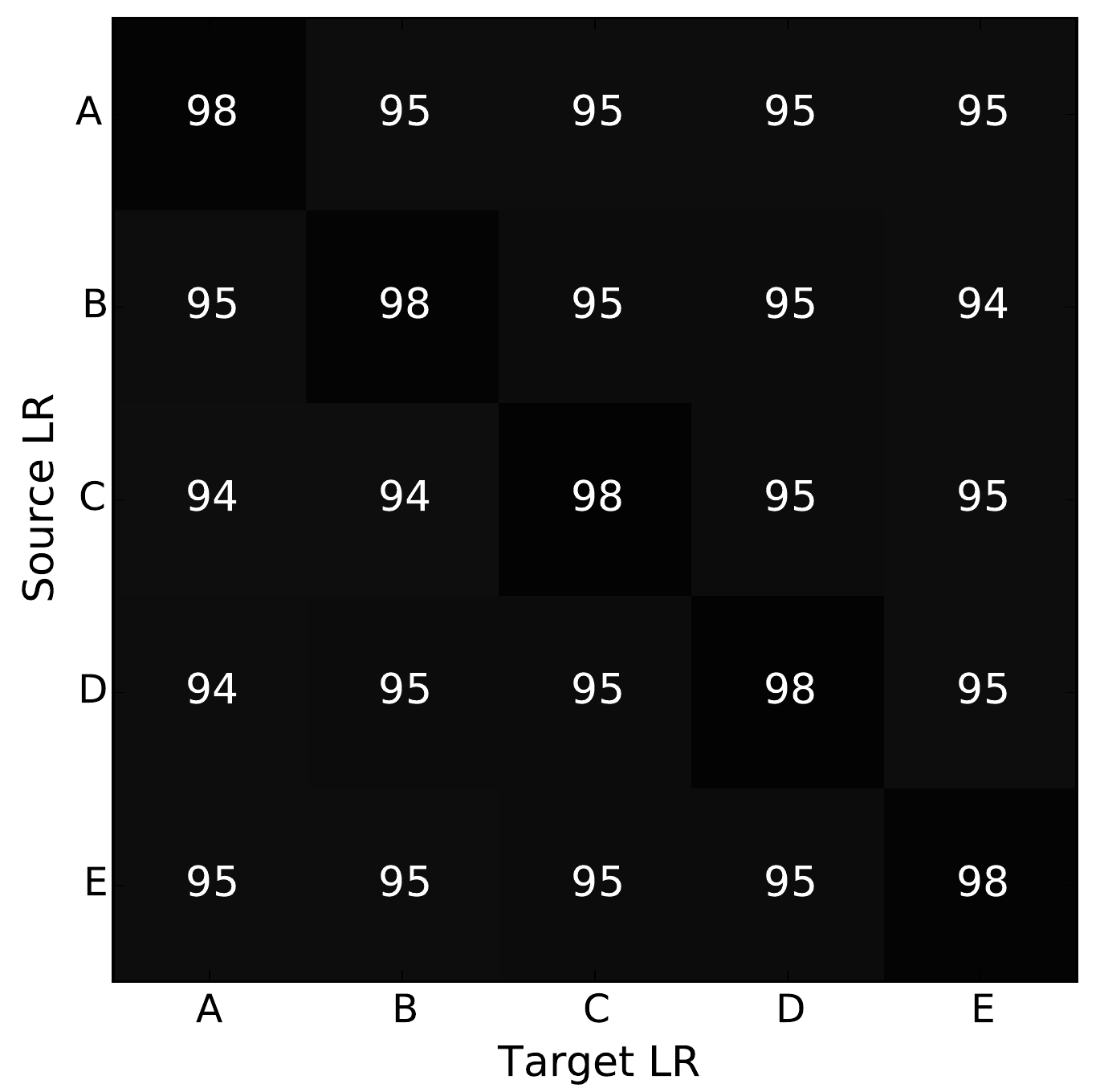} 
    \caption{LR models} 
    \label{fig:intramodel-transferability:c} 
    \vspace{4ex}
  \end{subfigure}
  \begin{subfigure}[b]{0.5\columnwidth}
    \centering
    \includegraphics[width=\columnwidth]{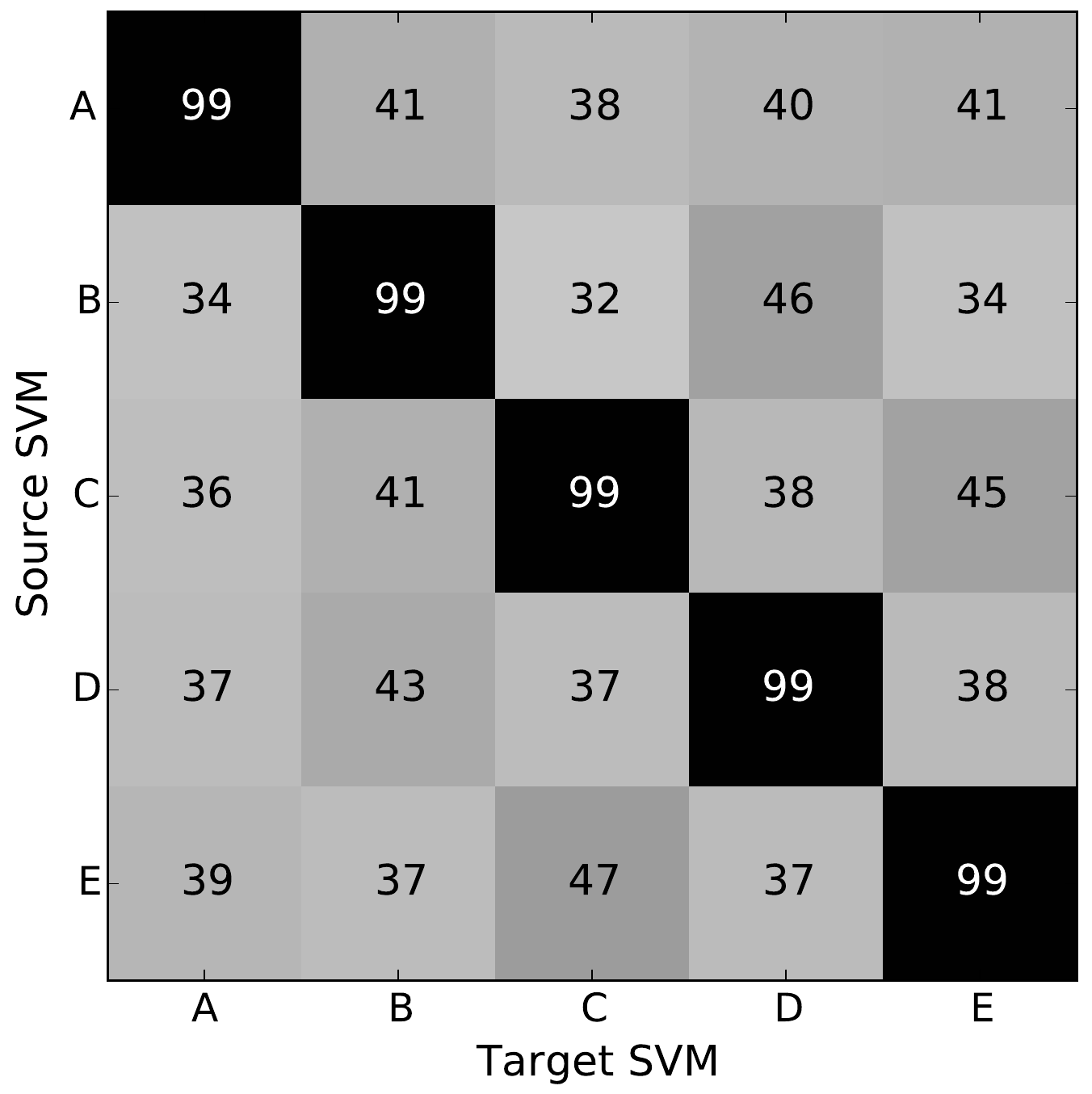} 
    \caption{SVM models} 
    \label{fig:intramodel-transferability:d} 
    \vspace{4ex}
  \end{subfigure}
  \begin{subfigure}[b]{0.5\columnwidth}
    \centering
    \includegraphics[width=\columnwidth]{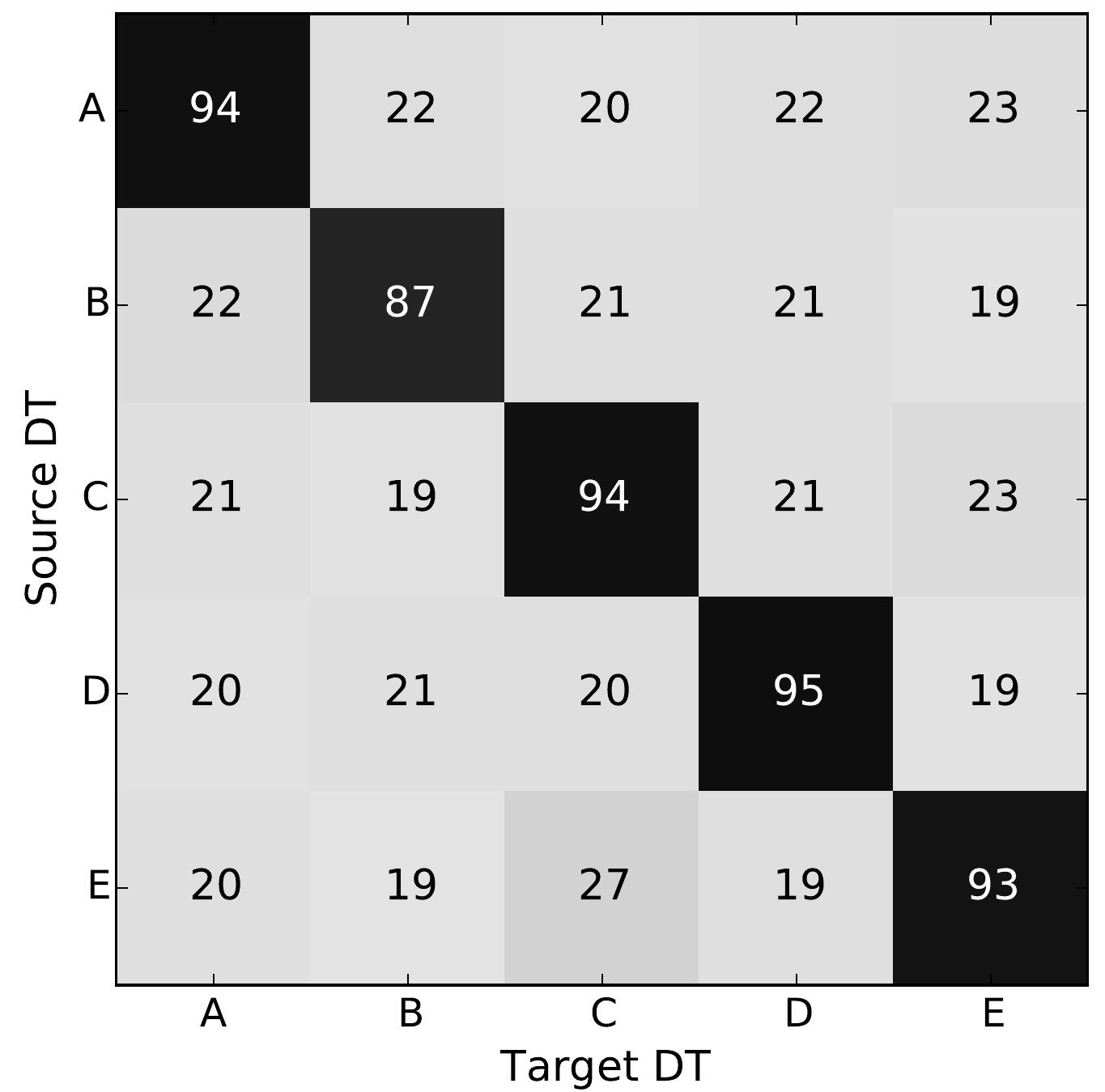} 
    \caption{DT models} 
    \label{fig:intramodel-transferability:e} 
  \end{subfigure}
	\begin{subfigure}[b]{0.5\columnwidth}
	  \centering
	  \includegraphics[width=\columnwidth]{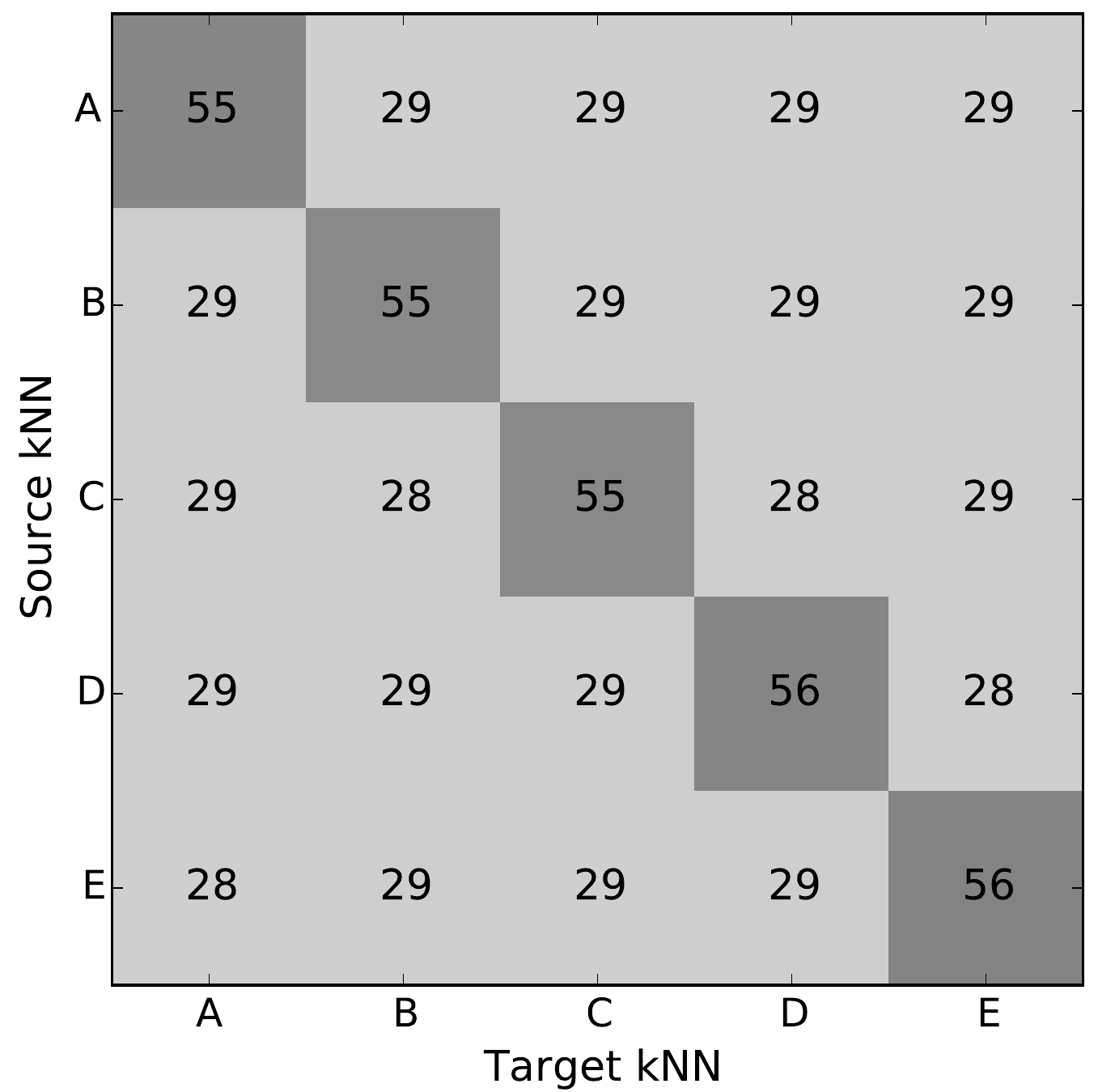} 
	  \caption{kNN models} 
	  \label{fig:intramodel-transferability:f} 
	\end{subfigure} 
  \caption{intra-technique transferability for 5 ML techniques. Figure~\ref{tbl:accuracies-intramodel} reports the accuracy rates of the 25 models used, computed on the MNIST test set. Figures~\ref{fig:intramodel-transferability:b}-\ref{fig:intramodel-transferability:f} are such that cell $(i,j)$ reports the intra-technique transferability between models $i$ and $j$, i.e. the percentage of adversarial samples produced using model $i$ misclassified by model $j$.}
  \label{fig:intramodel-transferability} 
\end{figure}

To train different models using the same machine learning technique, we split
the training set in disjoint subsets A,B,C,D,E of $10,000$ samples each, in
order of increasing indices. For each of the machine learning techniques (DNN,
LR, SVM, DT, kNN), we thus learn five different models referred to as
A,B,C,D,E. Model accuracies, i.e. the proportion of labels correctly predicted
by the model for the testing data, are reported in
Figure~\ref{tbl:accuracies-intramodel}. For each of the 25 models, we apply the
suitable adversarial sample algorithm described in
Section~\ref{sec:intra-technique-transferability} and craft $10,000$ samples from
the test set, which was unused during training. For adversarial sample
algorithms with parameters, we fine-tune them to achieve a quasi-complete
misclassification of the $10,000$ adversarial samples by the model on which
they are crafted. Upon empirically exploring the input variation parameter
space, we set it to $\varepsilon=0.3$ for the fast gradient sign method
algorithm, and $\varepsilon=1.5$ for the SVM algorithm.

Figures~\ref{fig:intramodel-transferability:b}-\ref{fig:intramodel-transferability:f}
report intra-technique transferability rates for each of the five machine learning
techniques. Rates $(i,i)$ on the diagonals indicate the proportion of
adversarial samples misclassified precisely by the same model $i$ on which they
were crafted. Off-diagonal rates $(i,j)$ indicate the proportion of adversarial
samples misclassified by a model $j$ different from the model $i$ on which they
were crafted. We first observe that all models are vulnerable to intra-technique
transferability in a non-negligible manner. LR models are most vulnerable as
adversarial samples transfer across models at rates larger than $94\%$. DNN
models display similarly important transferability, with  rates of at least
$49\%$. On the SVM, DT, and kNN matrices, the diagonals stand out more,
indicating that these techniques are to some extent more robust to the
phenomenon. In the case of SVMs, this could be explained by the explicit
constraint during training on the choice of hyperplane decision boundaries that
maximize the margins (i.e. support vectors). The robustness of both DTs and
kNNs could simply stem from their non-differentiability.

\subsection{Cross-technique Transferability}
\label{sec:transferability}

We define \emph{cross-technique transferability} between models $i$ and $j$,
 trained using different machine learning techniques, as the proportion of
adversarial samples produced to be misclassified by model $i$ that are also
misclassified by model $j$. Hence, this is a more complex phenomenon than
intra-technique transferability because it involves models learned using possibly
very different techniques like DNNs and DTs. Yet, cross-technique transferability
is surprisingly a strong phenomenon to which techniques like LR, SVM, DT, and
ensembles are vulnerable, making it easy for adversaries to craft adversarial
samples misclassified by models trained using diverse machine learning
techniques. 

We study the cross-technique transferability phenomenon across models trained using
the five machine learning techniques already used in
Section~\ref{sec:intra-technique-transferability} and described in
Section~\ref{sec:transferability-experimental-setup}
and~\ref{sec:adv-sample-crafting}. To these, we add a 6th model: an ensemble
$f(\vec{x})$. The ensemble $f$ is implemented using a collection of 5 experts,
which are the 5 previously described models: the DNN denoted $f_1$, LR denoted
$f_2$, SVM denoted $f_3$, DT denoted $f_4$, and kNN denoted $f_5$. Each expert
 makes a decision and the ensemble outputs the most frequent
choice (or the class with the lowest index if they all disagree):
\begin{equation}
f(\vec{x}) = \arg\max_{i\in 0..N-1} \sum_{j\in 1..5} f_{j,i}(\vec{x})
\end{equation}
where $f_{j,i}(\vec{x})=1_{f_j(\vec{x})==i}$ indicates whether classifier $f_j$ assigned class $i$ to input $\vec{x}$. Note that in this section, we only train one model per machine learning technique on the full MNIST training set of $50,000$ samples, unlike in Section~\ref{sec:intra-technique-transferability}. 

In this experiment, we are interested in transferability across machine
learning techniques. As such, to ensure our results are comparable, we
fine-tune the parameterizable crafting algorithms to produce adversarial
samples with similar perturbation magnitudes. To compare magnitudes across
perturbation styles, we use the L1 norm: the sum of each perturbation
component's absolute value. Perturbation added to craft adversarial samples
using the DNN, LR, and SVM have an average L1 norm $\|\delta_{\vec{x}}\|_1$ of
$11.5\%$. To achieve this, we use an input variation parameter of
$\varepsilon=0.25$ with the fast gradient sign method on the DNN, LR, and kNN.
To craft adversarial samples on the SVM, we use an input variation parameter of
$\varepsilon=5$ with the crafting method introduced in
Section~\ref{sec:adv-sample-crafting}. Unfortunately, the attack on DT cannot
be parameterized to match the L1 norm of DNN, LR, kNN and SVM attacks. Hence,
perturbations selected have much lower average L1 norms of respectively
$1.05\%$.

\begin{figure}[t] 
	\centering
	\includegraphics[width=0.9\columnwidth]{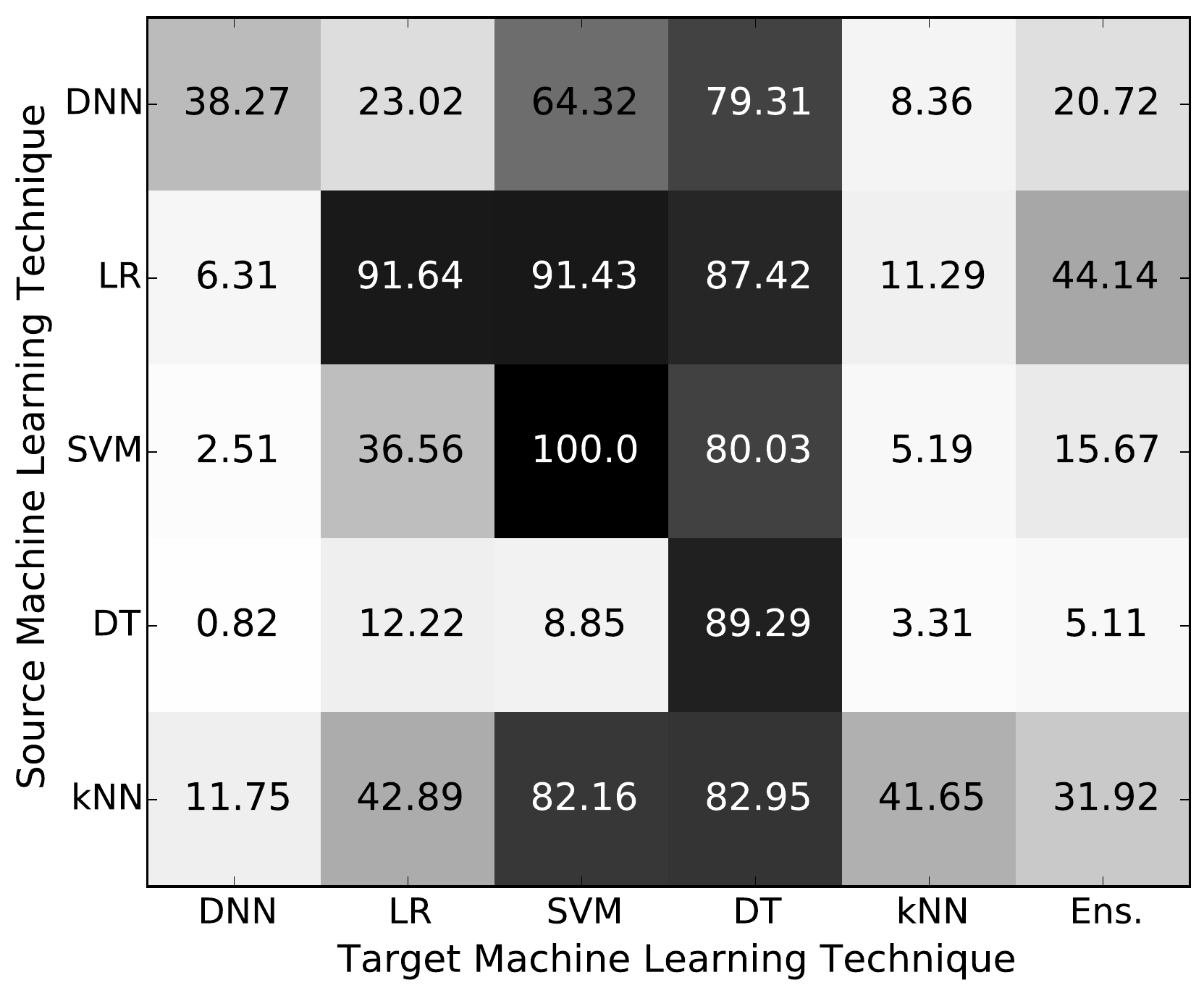} 
	\caption{cross-technique Transferability matrix: cell $(i,j)$ is the percentage of adversarial samples crafted to mislead a classifier learned using machine learning technique $i$ that are misclassified by a classifier trained with technique $j$.}
	\label{tbl:transferability-matrix}
\end{figure}

We build a cross-technique transferability matrix where each cell $(i,j)$ holds the
percentage of adversarial samples produced for classifier $i$ that are
misclassified by classifier $j$. In other words, rows indicate the machine
learning technique that trained the model against which adversarial samples
were crafted. The row that would correspond to the ensemble is not included
because there is no crafting algorithm designed to produce adversarial samples
specifically for an ensemble, although we address this limitation in
Section~\ref{sec:learning-approximators} using insight gained in this
experiment. Columns indicate the underlying technique of the classifier making
predictions on adversarial samples. This matrix, plotted in
Figure~\ref{tbl:transferability-matrix}, shows that cross-technique transferability
is a strong but heterogeneous phenomenon. The most vulnerable model is the
decision tree (DT) with misclassification rates ranging from $47.20\%$ to
$89.29\%$ while the most resilient is the deep neural network (DNN) with
misclassification rates between $0.82\%$ and $38.27\%$. Interestingly, the
ensemble is not resilient to cross-technique transferability of adversarial samples
with rates reaching $44.14\%$ for samples crafted using the LR model. This is
most likely due to the vulnerability of each underlying expert to adversarial
samples.

We showed that all machine learning techniques we studied are vulnerable to two types of adversarial sample
transferability. This most surprisingly results in adversarial samples being
misclassified across multiple models learned with different machine learning techniques. This \emph{cross-technique transferability} greatly reduces the minimum knowledge that adversaries must
possess of a machine learning classifier in order to force it to misclassify inputs that they crafted. We leverage this
observation, along with findings from
Section~\ref{sec:learning-approximators}, to justify design choices in the
attack described in Section~\ref{sec:ml-oracle-attack}.


\section{Learning Classifier Substitutes by Knowledge Transfer}
\label{sec:learning-approximators}

In the previous section, we identified machine learning techniques (e.g., DNNs and LR) yielding
models adequate for crafting samples  misclassified across models
trained with different techniques, i.e adversarial samples
with strong cross-technique transferability. Thus, in order to craft adversarial 
samples misclassified by a classifier whose underlying model is unknown,
adversaries can instead use a \emph{substitute} model if it solves the same classification problem and its parameters are known. Therefore, efficiently learning substitutes is 
key to designing \emph{black-box attacks} where adversaries target remote
classifiers whose model, parameters, and training data are unknown to them.  This is precisely the attack
scenario evaluated against commercial machine learning platforms in
Section~\ref{sec:ml-oracle-attack}, while we focus in this section on the
prerequisite learning of substitutes for machine learning classifiers.

We enhance an  algorithm introduced in~\cite{papernot2016practical}
to learn a substitute model for a given
classifier simply by querying it for labels on carefully chosen inputs. More precisely, we introduce two refinements to the algorithm: one improves its accuracy and the second reduces its computational complexity. 
We generalize the learning of
substitutes to oracles using a range of machine learning techniques: DNNs, LR, SVMs, DTs, and kNNs.
Furthermore, we show that both DNNs and LR can be used as substitute models for all machine learning techniques studied to the exception of decision trees.

\subsection{Dataset Augmentation for Substitutes}
\label{sec:jacobian-based-dataset-augmentation}

The targeted
classifier is designated as an \emph{oracle} because adversaries have the
minimal capability of querying it for predictions on inputs of
their choice. The oracle returns the \emph{label} (\textbf{not} the probabilities) assigned to the 
sample.
No other knowledge of the
classifier (e.g., model type, parameters, training data) is available. To
circumvent this, we build on a technique introduced
in~\cite{papernot2016practical}, which leverages a dataset augmentation
technique to train the substitute model.  

\textbf{Jacobian-based dataset augmentation -} We use this augmentation
technique introduced in~\cite{papernot2016practical} to learn DNN and LR substitutes for oracles. First, one collects
an initial substitute training set of limited size (representative of the task
solved by the oracle) and labels it by querying the oracle. Using this labeled
data, we train a first substitute model $f$ likely to perform poorly as a
source of adversarial samples due to the small numbers of samples used for
training. To select additional training points, we use the following:
\begin{equation}
\label{eq:jacobian-based-heuristic}
S_{\rho+1} = \{ \vec{x}+\lambda_\rho \cdot \sgn(J_f[\tilde{O}(\vec{x})]:\vec{x}\in S_\rho)  \} \cup S_\rho
\end{equation}
where $S_\rho$ and $S_{\rho+1}$ are the previous and new training sets,
$\lambda_\rho$ a parameter fine-tuning the augmentation step size, $J_f$ the
Jacobian matrix of  substitute $f$, and $\tilde{O}(\vec{x})$ the oracle's label
for sample $\vec{x}$. We train a new instance $f$ of the substitute with the
augmented training set $S_{\rho+1}$, which we can label simply by querying
oracle $\tilde{O}$. By alternatively augmenting the training set and training a
new instance of the substitute model for multiple iterations $\rho$, Papernot
et al. showed that substitute DNNs can approximate another
DNNs~\cite{papernot2016practical}.  

\textbf{Periodical Step Size -} When introducing the technique, Papernot et al. used a fixed step size parameter
$\lambda_\rho$ throughout the substitute learning iterations $\rho$. In this section, we
show that by having a step size periodically alternating between positive and negative
values, one can improve the quality of the oracle approximation made by the substitute, which we measure in terms
of the number of labels matched with the
original classifier oracle. More precisely, we introduce an iteration period $\tau$
after which the step size is multiplied by $-1$. Thus, the step size
$\lambda_\rho$ is defined as:
\begin{equation}
\label{fig:periodical-step-size}
\lambda_\rho = \lambda \cdot (-1)^{\left\lfloor \frac{\rho}{\tau} \right\rfloor}
\end{equation}
where $\tau$ is set to be the number of epochs after which the Jacobian-based
dataset augmentation does not lead any substantial improvement in the
substitute. A grid search can also be performed to find an optimal value for
the period $\tau$. We also experimented with a decreasing grid step amplitude
$\lambda$, but did not find that it yielded substantial improvements.

\textbf{Reservoir Sampling -} We also introduce the use of \emph{reservoir sampling}~\cite{vitter1985random}
as a mean to reduce the number of queries made to the oracle. This is useful
when learning substitutes in realistic environments where the number of label queries an adversary can make without
exceeding a quota or being detected by a defender is
constrained. Reservoir
sampling is a class of algorithms that randomly select $\kappa$ samples from a list of samples. The
total number of samples in the list can be both very large and unknown. In our
case, we use reservoir sampling to select a limited number of new inputs
$\kappa$ when performing a Jacobian-based dataset augmentation. This prevents
the exponential growth of queries made to the oracle at each augmentation
iteration. At iterations $\rho>\sigma$ (the first $\sigma$ iterations are
performed normally), when considering the previous set $S_{\rho-1}$ of
substitute training inputs, we select $\kappa$ inputs from $S_{\rho-1}$ to be
augmented in $S_\rho$. These $\kappa$ inputs are selected using reservoir
sampling, as described in Algorithm~\ref{alg:reservoir-sampling}. This
technique ensures that each input in $S_{\rho-1}$ has an equal probability
$\frac{1}{\left| S_{\rho-1} \right|}$ to be augmented in $S_\rho$. The number
of queries made to the oracle is reduced from $n\cdot2^\rho$ for the vanilla
Jacobian-based augmentation to $n\cdot2^\sigma+\kappa\cdot(\rho-\sigma)$ for
the Jacobian-based augmentation with reservoir sampling. Our experiments show
that the reduced number of training points in the reservoir sampling variant
does not significantly degrade the quality of the substitute.

\begin{algorithm}[t]
\caption{Jacobian-based augmentation with Reservoir Sampling: sets are considered as arrays for ease of notation.}
\label{alg:reservoir-sampling}
\begin{algorithmic}[1]
	\Require $S_{\rho-1}$, $\kappa$, $J_f$, $\lambda_\rho$
	\State $N \leftarrow \left| S_{\rho-1} \right|$
	\State Initialize $S_{\rho}$ as array of $N+\kappa$ items
	\State $S_{\rho}[0:N-1] \leftarrow S_{\rho-1}$
	\For{$i\in 0..\kappa-1$}
		\State $S_{\rho}[N+i] \leftarrow S_{\rho-1}[i]+\lambda_\rho \cdot \sgn(J_f[\tilde{O}(S_{\rho-1}[i])]) $
	\EndFor
	\For{$i\in \kappa .. N-1$}
		\State $r\leftarrow$ random integer between $0$ and $i$
		\If{$r < \kappa$}
			\State $S_{\rho}[N+r] \leftarrow S_{\rho-1}[i]+\lambda_\rho \cdot \sgn(J_f[\tilde{O}(S_{\rho-1}[i])]) $
		\EndIf
	\EndFor
	\State \Return $S_{\rho}$
\end{algorithmic}
\end {algorithm}

\subsection{Deep Neural Network Substitutes} 
\label{sec:dnn-approximators}

In~\cite{papernot2016practical}, the oracle classifier approximated was always
a DNN. However, the authors concluded with preliminary results suggesting
applicability to a nearest neighbors classifier. We here show that in fact the
technique is generalizable and applicable to many machine learning techniques
by evaluating its performance on 5 types of ML classifiers: a DNN, LR, SVM, DT,
and kNN. This spectrum is representative of machine learning (cf.
Section~\ref{sec:transferability-experimental-setup}). Our experiments suggest
that one can accurately \emph{transfer} the knowledge from \emph{many} machine learning
classifiers to a DNN and obtain a DNN mimicking the decision boundaries of the
original classifier.

Using the Jacobian-based augmentation technique, we train 5 different
substitute DNNs to match the labels produced by 5 different oracles, one for
each of the ML techniques mentioned.
These classifiers serving as oracles are all trained on the $50,000$ sample
MNIST training set using the models described previously in
Section~\ref{sec:transferability-experimental-setup}. To approximate them, we
use the first $100$ samples from the MNIST test set (unseen during training) as
the initial substitute training set and follow three variants of the procedure
detailed in Section~\ref{sec:jacobian-based-dataset-augmentation} with
$\lambda=0.1$: (1) vanilla Jacobian-based augmentation, (2) with $\tau=3$
periodic step size, (3) with both $\tau=3$ periodic step size and reservoir
sampling with parameters $\sigma=3$ and $\kappa=400$. The substitute
architecture is identical to the DNN architecture from
Section~\ref{sec:transferability-experimental-setup}. We allow experiments to
train substitutes for 10 augmentation iterations, i.e. $\rho\leq 9$.

\begin{figure}[t!] 
	\begin{subfigure}[b]{\columnwidth}
	  \centering
	  \includegraphics[width=0.95\columnwidth]{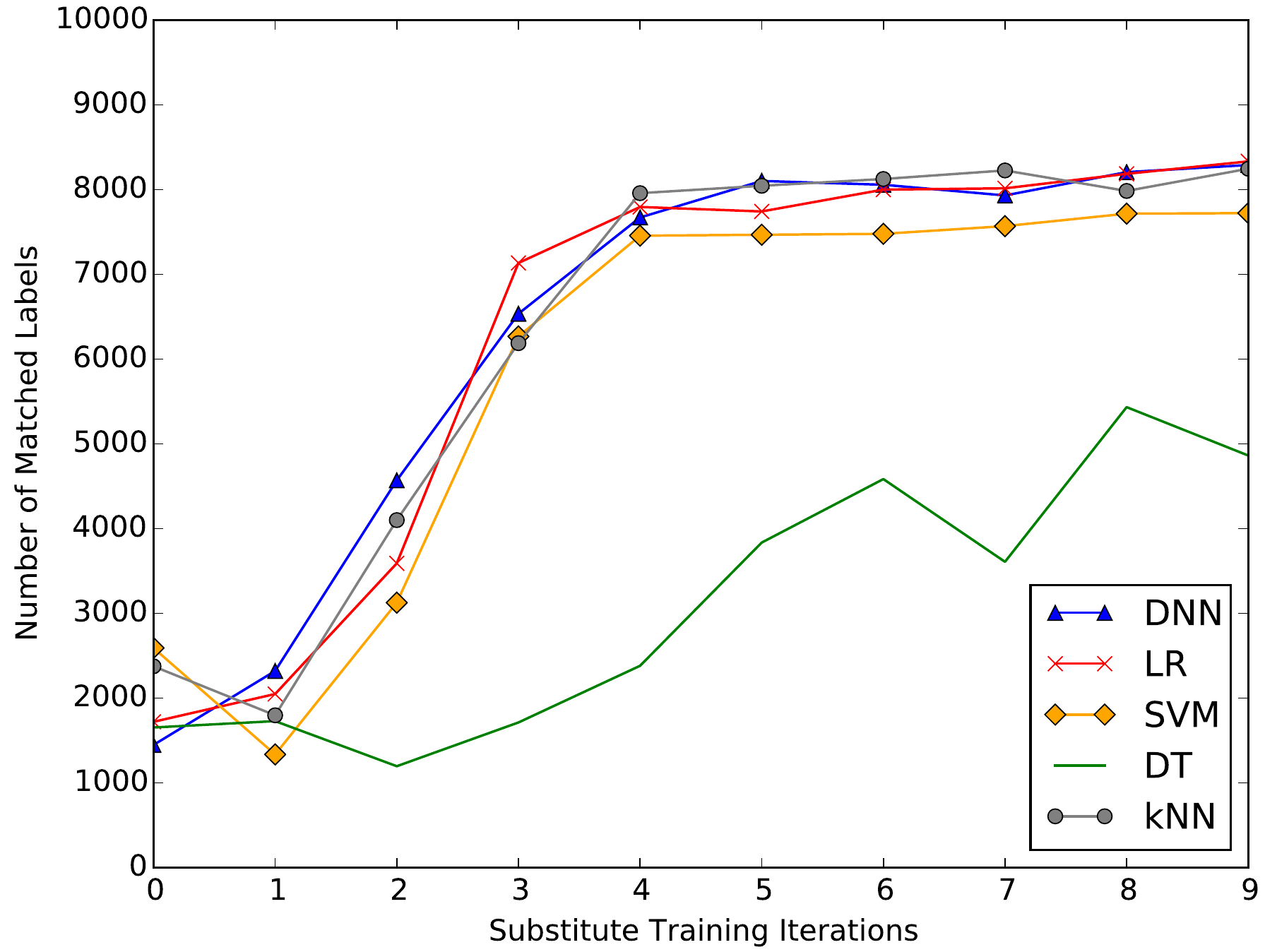} 
	  \caption{DNN substitutes} 
	  \label{fig:learning-approximators-dnn} 
	  \end{subfigure}
	  	\begin{subfigure}[b]{\columnwidth}
	  	  \centering
	  	  \includegraphics[width=0.95\columnwidth]{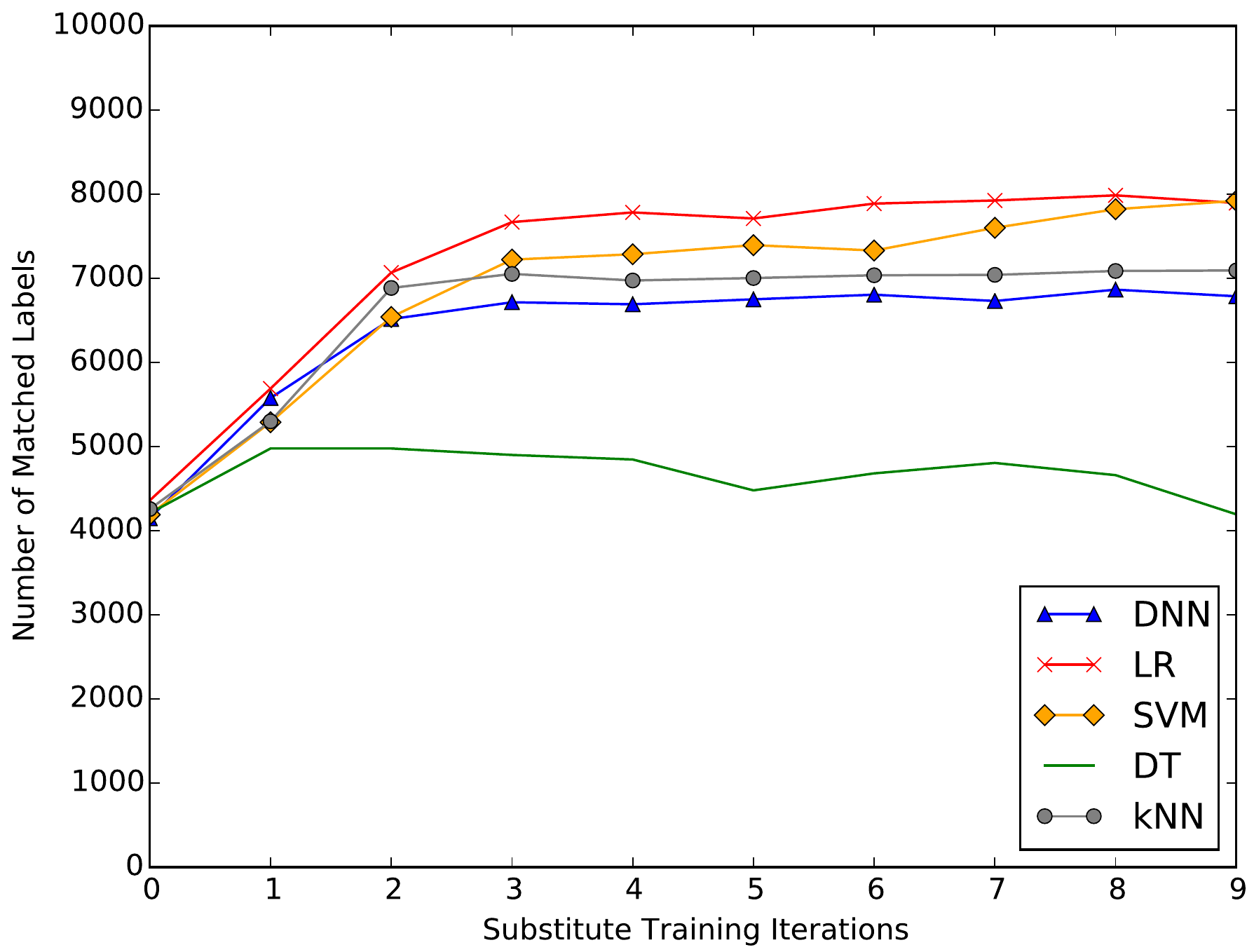} 
	  	  \caption{LR substitutes} 
	  	  \label{fig:learning-approximators-lr} 
	  	  \end{subfigure}
	\caption{Label predictions matched between the DNN and LR substitutes and their target classifier oracles on test data.}
\end{figure}

Figure~\ref{fig:learning-approximators-dnn}  plots at each iteration $\rho$ the
share of samples on which the substitute DNNs agree with predictions made by
the classifier oracle they are approximating.  This proportion is estimated by
comparing the labels assigned to the MNIST test set by the substitutes and
oracles before each iteration $\rho$ of the Jacobian-based dataset
augmentation. The substitutes used in this figure were all trained with both
a periodic step size and reservoir sampling, as described previously. 
Generally speaking, all substitutes are able to successfully
approximate the corresponding oracle, after $\rho=10$ augmentation iterations,
the labels assigned match for about $77\%$ to $83\%$ of the MNIST test set,
except for the case of the DT oracle, which is only matched for $48\%$ of the samples. This difference could be explained by the
non-differentiability of decisions trees. On the contrary, substitute DNNs are
able to approximate the nearest neighbors oracle although it uses lazy
classification: no model is learned at training time and predictions are made
by finding close training sample(s).

The first three rows of Table~\ref{tbl:refinements} quantify the impact of
the two refinements introduced above on the proportion of test set labels
 produced by the oracle that were matched by DNN substitutes. 
The first refinement, the periodic step size, allows
substitutes to approximate more accurately their target oracle. For instance at
$\rho=9$ iterations, the substitute DNN trained with a periodic ste size
for the DNN oracle matches $89.28\%$ of the labels whereas the vanilla substitute 
DNN only matched $78.01\%$. Similarly, the substitute DNN trained with a periodic ste size
for the SVM oracle matches $83.79\%$ of the labels whereas the vanilla substitute 
 only matched $79.68\%$.
The second refinement, reservoir sampling allows us to train substitutes for
more augmentation iterations without making too many queries to the oracle. For
instance, $10$ iterations with reservoir sampling (using $\sigma=3$ and $\kappa
=400$) make $100\cdot 2^3 + 400(10-3)=3,600$ queries to the oracle instead of
$102,400$ queries with the vanilla technique. The reduced number of queries has
an impact on the substitute quality compared to the periodic step size
substitutes but it is still superior to the vanilla substitutes. For instance,
when approximating a DNN oracle, the vanilla substitute matched $7,801$ labels,
the periodic step size one $8,928$, and the periodic step size with reservoir
sampling one $8,290$.

\begin {table}[t!]
\centering
\begin{tabular}{|l||c|c|c|c|c|}
	\hline
	Substitute  & DNN & LR & SVM & DT & kNN     \\  \hline \hline
	DNN & 78.01  & 82.17  & 79.68  & 62.75  & 81.83  \\ \hline
	DNN+PSS & 89.28  & 89.16  & 83.79   & 61.10  & 85.67  \\ \hline
	DNN+PSS+RS & 82.90  & 83.33   & 77.22  & 48.62  & 82.46   \\ \hline\hline
	LR & 64.93  & 72.00  & 71.56  & 38.44  & 70.74   \\ \hline
	LR+PSS & 69.20 & 84.01  & 82.19  & 34.14  & 71.02 \\ \hline
	LR+PSS+RS & 67.85  & 78.94  & 79.20  & 41.93  & 70.92  \\ \hline
	 \hline
\end{tabular}
\caption{Impact of our refinements, Periodic Step Size (PSS) and Reservoir Sampling (RS), on the percentage of label predictions matched between the substitutes and their target classifiers on test data  after $\rho=9$ substitute iterations. }
\label{tbl:refinements}
\end{table}

\subsection{Logistic Regression Substitutes}
\label{sec:lr-approximators}

Having generalized substitute learning with a demonstration of the
capacity of DNNs to approximate any machine learning model, we now
consider replacing the substitute itself by another machine learning technique.
Experiments in Section~\ref{sec:transferability} led us to conclude that cross-technique transferability is not specific to adversarial
samples crafted on DNNs, but instead applies to many
learning techniques. Looking at
Figure~\ref{tbl:transferability-matrix} again, a natural candidate is logistic
regression, as it displays large cross-technique transferability rates superior to
DNNs except when targeting DNNs themselves.

The Jacobian-based dataset augmentation's implementation for DNNs is easily
adapted to multi-class logistic regression. Indeed, multi-class logistic
regression is analog to the softmax layer frequently used by deep neural
networks to produce class probability vectors. We can easily compute the
$(i,j)$ component of the Jacobian of a multi-class LR model:
\begin{equation}
\label{lr-jacobian}
J_f(\vec{x}) [i,j] =  \frac{w_j e^{\vec{w_j}[i]\cdot\vec{x}}-\sum_{l=1}^N \vec{w_l}[i]\cdot e^{w_l \vec{x}}}{\left(\sum_{l=1}^N e^{\vec{w_l}[i]\cdot \vec{x}}\right)^2}
\end{equation}
where notations are the ones used in Equation~\ref{eq:logistic-regression}. 

Hence, we repeat the experiment from Section~\ref{sec:dnn-approximators} but we
now train multi-class logistic regression substitute models (instead of the DNN
substitutes) to match the labels produced by the classifier oracles. Everything
else is unchanged in the experimental setup. As illustrated in Figure~\ref{fig:learning-approximators-lr}, the change of model type for the
substitute generally speaking degrades the approximation quality: the
proportion of labels matched is reduced. Performances of LR substitutes are
competitive with those of DNN substitutes for LR and SVM oracles. Here again,
the substitutes perform poorly on the decision tree oracle, with match rates
barely above $40\%$. 

The last three rows of Table~\ref{tbl:refinements} quantify the impact of
the two refinements introduced above on the proportion of test set labels
produced by the oracle that were matched by LR substitutes. 
The first refinement, the periodic step size, allows LR
substitutes to approximate more accurately their target oracle, as was also the 
case for DNN substitutes. For instance at
$\rho=9$ iterations, the LRsubstitute trained with a periodic ste size
for the LR oracle matches $84.01\%$ of the labels whereas the vanilla LR substitute  only matched $72.00\%$. Similarly, the LR substitute trained with a periodic ste size
for the SVM oracle matches $82.19\%$ of the labels whereas the vanilla substitute 
only matched $71.56\%$.
The second refinement, reservoir sampling allows us to reduce the number of queries
with a limited impact on the substitute quality: less labels are match than 
the periodic step size substitutes but more than the vanilla substitutes. 
For instance,
when approximating a SVM oracle, the vanilla substitute matched $71.56\%$ of the labels, the periodic step size one $82.19\%$, and the periodic step size with reservoir sampling one $79.20\%$.

The benefit of vanilla LR substitutes compared
to DNN substitutes is that they achieve their asymptotic match rate faster,
after only $\rho=4$ augmentation iterations, corresponding to $1,600$ oracle
queries. Furthermore, LR models are much lighter in terms of computational
cost. These two factors could justify the use of LR (instead of DNN)substitutes
in some contexts. The reservoir sampling technique gives good performances,
especially on LR and SVM oracles.

\subsection{Support Vector Machines Substitutes}

Having observed that deep learning and logistic regression were both relevant
when approximating classifier oracles, we now 
turn to SVMs for substitute learning. This is motivated by the strong cross-technique transferability of adversarial sample crafted
using an SVM observed in Section~\ref{sec:transferability-section}, making SVMs good candidates for substitutes in a black-box attack.

\textbf{SVM-based dataset augmentation -} To train SVMs to approximate
oracles in a manner analogous to the Jacobian-based dataset augmentation, we introduce a new augmentation
technique. We
replace the heuristic in Equation~\ref{eq:jacobian-based-heuristic}
by the following, which is adapted to the specificities of SVMs: 
\begin{equation}
\label{eq:svm-dataset-augmentation}
S_{\rho+1} = \{ \vec{x} - \lambda \cdot \frac{\vec{w}[\tilde{O}(\vec{x})]}{\left\| \vec{w}[\tilde{O}(\vec{x})] \right\|}\vec{x}:\vec{x}\in S_\rho)  \} \cup S_\rho
\end{equation}
where $\vec{w}[k]$ is the weight indicating the hyperplane direction of
subclassifier $k$ used to implement a multi-class SVM with the one-vs-the-rest
scheme as detailed in Equation~\ref{eq:sub-svm-binary}. This heuristic selects
new points in the direction orthogonal to the hyperplane acting as the decision
boundary for the binary SVM subclassifier $k$ corresponding to the input's
label. This is precisely the direction used in Equation~\ref{eq:svm-adv-sample}
to find adversarial samples but parameter $\lambda$ is here generally set to
lower values so as to find samples \emph{near} the decision boundary instead of on
the other side of the decision boundary. 

\textbf{Experimental Validation -} We
repeat the experiments from Sections~\ref{sec:dnn-approximators}
and~\ref{sec:lr-approximators} but we now train 18 different SVM
models to match labels produced by the classifiers---instead of training DNN or LR substitutes. Unfortunately, our results suggest that
SVMs are unable to perform knowledge transfer from oracles that are not SVMs
themselves using the dataset augmentation technique introduced in Equation~\ref{eq:svm-dataset-augmentation}, as well as the refinements introduced previously: the periodic step size and reservoir sampling. Indeed, the SVM substitute matches $79.80\%$ of the SVM oracle labels, but only $11.98\%$ and $11.97\%$ of the DNN and LR oracle labels. These numbers are not improved by the use of a periodic step size and/or reservoir sampling. This could be due to the specificity of SVM training and the decision boundaries they learn. Future work should investigate the use of alternative augmentation techniques to confirm our findings.

In this section, we evaluated the capacity of DNN, LR, and SVM substitutes
 to approximate a classifier oracle by querying it for labels  on inputs
selected using a heuristic relying on the substitute's Jacobian. We observed that predictions
made by DNN and LR substitutes more accurately matched the
targeted oracles than SVM substitute predictions. We
emphasize that all experiments only required knowledge of $100$
samples from the MNIST test set. In other words, learning substitutes does not
require knowledge of the targeted classifier's type, 
parameters, or training data, and can thus be performed under realistic adversarial threat
models. 


\section{Black-Box Attacks of Remote\\ Machine Learning Classifiers}
\label{sec:ml-oracle-attack}

Intra-technique and cross-technique transferability of adversarial samples, together with
the learning of substitutes for classifier oracles, enable a range of attacks
targeting remote machine learning based systems whose internals are unknown to adversaries. To
illustrate the feasibility of \emph{black-box attacks} on such remote systems,
we target in an experiment two machine learning classifiers respectively trained and hosted
by Amazon and Google. We find it is possible to craft samples misclassified by
these commerical oracles at respective rates of $96.19\%$ and
$88.94\%$ after making $800$ queries to learn substitute models approximating them.

\subsection{The Oracle Attack Method}

This section's adversarial threat model is identical to the one used when
learning substitutes in Section~\ref{sec:learning-approximators}: adversaries
have an \emph{oracle} access to the remote classifier. Its type, parameters, or
training set are all unknown to the adversary. The attack method leverages
Sections~\ref{sec:transferability-section}
and~\ref{sec:learning-approximators} of this paper, and is a generalization of
the approach introduced in~\cite{papernot2016practical}. 

The adversary first locally trains a substitute model to
approximate the remotely hosted classifier, using queries to the oracle as described in
Section~\ref{sec:learning-approximators}. We consider the use of deep learning
and logistic regression to learn substitutes for classifiers. We apply
the two refinements introduced in this paper: a periodic step size
and reservoir sampling. Since substitute models are locally trained, the
adversary has full knowledge of their model parameters. Thus, one of
the adversarial sample crafting algorithms introduced in
Section~\ref{sec:adv-sample-crafting} corresponding to the machine learning
technique used to learn the substitute are employed to craft adversarial samples
misclassified by the substitute model. The adversary than leverages either intra-technique
or cross-technique transferability of adversarial samples---depending on the techniques with which the substitute and oracle were learned: the inputs misleading
the locally trained substitute model are very likely to also deceive the targeted remotely
hosted oracle.  

Previous work conducted such an attack using a substitute and targeted
classifier both trained using deep learning, demonstrating that the attack was
realistic using the MetaMind API providing Deep Learning as a
Service~\cite{papernot2016practical}. We generalize these results by performing
the attack on Machine Learning as a Service  platforms that employ techniques
that are unknown to us: Amazon Web Services and Google Cloud Prediction. Both
platforms automate the process of learning classifiers using a labeled dataset
uploaded by the user. Unlike MetaMind, neither of these platforms claim to
exclusively use  deep learning to build classifiers. When analyzing our
results, we found that Amazon uses logistic regression (cf. below) but to the
best of our knowledge Google has never disclosed the technique they use to
train classifiers, ensuring that our experiment is properly blind-folded.

\subsection{Amazon Web Services Oracle}

Amazon offers a machine learning service, \emph{Amazon Machine
Learning},\footnote{\url{https://aws.amazon.com/machine-learning}} as part of
their Amazon Web Services platform. We used this service to train and host a ML
classifier oracle. First, we uploaded a CSV encoded version of the MNIST
training set to an S3 bucket on Amazon Web Services. We truncated the pixel
values in the CSV file to $8$ decimal places. We then started the ML model
training process on the Machine Learning service: we loaded the CSV training
data from our S3 bucket, selected the multi-class model type, provided the
target column in the CSV file, and kept the default configuration settings.
Note that Amazon offers limited customization options: the settings allow one
to customize the recipe (data transformations), specify a maximum model size
and number of training epochs, disable training data shuffle, and change the
regularization type between L1 and L2 or simply disable regularization. The
training process takes a few minutes and outputs a classifier model achieving a
$92.17\%$ accuracy on the MNIST test set. We have no way to improve that
performance beyond the limited customizing options as the intent of the service
is to automate model training. Finally, we activate real-time predictions to be
able to query the model for labels from our local machine.

We then use the Python API provided with the Amazon Machine Learning service to
submit prediction queries to our trained oracle model and retrieve the output
label. Although confidence values are available for predictions, we only
consider the label to ensure our threat model for adversarial capabilities
remains realistic. We incorporate this oracle in our experimental setup and
train two substitute models to approximate the labels produced by this oracle,
a DNN and LR, as SVM substitutes were dismissed by the conclusions of Section~\ref{sec:learning-approximators}. We train two variants of the DNN and LR substitutes. The first variant is trained with the vanilla dataset augmentation and the second variant with the enhanced dataset augmentation introduced in this paper, which uses both a periodic step size and reservoir sampling. 
Learning is initialized
with a substitute training set of $100$ samples
from the MNIST test set. For all substitutes, we measure the attack success  as the proportion among the $10,000$ adversarial
samples, produced using the fast gradient sign method with parameter
$\varepsilon=0.3$ (cf. Section~\ref{sec:adv-sample-crafting}) and the MNIST test set, misclassified by the Amazon oracle.

\begin {table}[t]
\centering
\begin{tabular}{|c||c|c|}
	\hline
	Substitute type  & DNN & LR      \\  \hline \hline
	$\rho=3$ (800 queries) & 87.44\% &  96.19\%   \\ \hline
	$\rho=6$ (6,400 queries)  & 96.78 \% & 96.43\%   \\ \hline\hline
	$\rho=6$  (PSS + RS) (2,000 queries) & 95.68\% & 95.83\%   \\ \hline
\end{tabular}
\caption{Misclassification rates of the Amazon oracle on adversarial samples ($\varepsilon=0.3$) produced with DNN and LR substitutes after $\rho=\{3,6\}$ augmentation iterations. Substitutes are trained without and with refinements from Section~\ref{sec:learning-approximators}: periodic step size (PSS) and reservoir sampling (RS).}
\label{tbl:aws-misclassification}
\end{table}

Misclassification rates of the Amazon Machine Learning oracle on adversarial
samples crafted using both the DNN and LR substitutes after $\rho\in\{3,6\}$ dataset augmentation iterations
are reported in Table~\ref{tbl:aws-misclassification}. Results are given for models learned without and with the two refinements---periodic step size (PSS) and reservoir sampling (RS)---introduced in Section~\ref{sec:learning-approximators}. With a misclassification
rate of $96.19\%$ for an adversarial perturbation $\varepsilon=0.3$ using a LR substitute
trained with $800$ queries ($\rho=3$) to the oracle, the model trained by Amazon is easily
misled. To understand why, we carefully read the online documentation and eventually found
one page indicating that the type of model trained by the Amazon Machine
Learning service is an ``industry-standard'' multinomial logistic
regression.\footnote{\url{http://docs.aws.amazon.com/machine-learning/latest/dg/types-of-ml-models.html}}
As seen in Section~\ref{sec:transferability-section}, LR is extremely vulnerable to
intra-technique and to a lesser extend vulnerable to cross-technique
transferability. In fact, as pointed out by Goodfellow et
al.~\cite{goodfellow2014explaining}, shallow models like logistic regression
are unable to cope with adversarial samples and learn a classifier resistant to
them. This explains why (1) the attack is very successful and (2) the LR
substitute performs better than the DNN substitute.

Additionally, Table~\ref{tbl:aws-misclassification} shows how the use of a periodic step size (PSS) together with
reservoir sampling (RS) allows us to reduce the number 
of queries made to the Amazon oracle while learning 
a DNN substitute producing adversarial samples with 
higher transferability to the targeted classifier. 
Indeed, we reduce by a factor of more than $3$ the 
number of queries made from $6,400$ to $2,000$, while only degrading the misclassification rate from $96.78\%$ to $95.68\%$---still larger than 
the rate of $87.44\%$ achieved after $800$ queries by the substitute learned without PSS and RS. For the LR substitutes, 
we do not see any positive impact from the use of PSS 
and RS, which is most likely to the fast convergence
of LR substitute learning, as observed in Section~\ref{sec:learning-approximators}.

\subsection{Google Cloud Prediction Oracle}

To test whether this poor performance is limited to the Amazon Web Services
platform, we now target the Google Cloud Prediction API service\footnote{\url{https://cloud.google.com/prediction/}}. The procedure
to train a classifier on Google's platform is similar to Amazon's. We first
upload to Google's Cloud Storage service the CSV encoded file of the MNIST
training data identical to the one used to train the oracle on Amazon Machine
Learning. We then activate the Prediction API on Google's Cloud Platform and
train a model using the  API's method named
\texttt{prediction.trainedmodels.insert}. The only property we are able to
specify is the expected multi-class nature of our classifier model as well as
the column in the CSV indicating target labels. We then evaluate the resulting
model using the API method \texttt{prediction.trainedmodels.predict} and an uploaded CSV file of the MNIST test set. The API  reports
an accuracy of $92\%$ on this test set for the model trained.

\vspace*{-0.05in}

We now use the Google Cloud Python API to connect our experimental setup to the Prediction API, thus allowing our algorithms to
make queries to the Google classifier oracle. As we did for Amazon, we train two
substitute models (DNN and LR) using an initial substitute training set of 100
samples from the MNIST test set. For each substitute type, we train two model variants: the first one without periodic step size (PSS) or reservoir sampling (RS), the second one with both PSS and RS. Table~\ref{tbl:google-misclassification}
reports the rate of adversarial samples produced by each of the four resulting substitutes
and misclassified by the Google Prediction API oracle. 

\begin {table}[t]
\centering
\begin{tabular}{|c||c|c|}
	\hline
	Substitute type  & DNN & LR      \\  \hline \hline
	$\rho=3$ (800 queries) & 84.50\% &  88.94\%   \\ \hline
	$\rho=6$ (6,400 queries)  & 97.17\% & 92.05\%   \\ \hline\hline
	$\rho=6$  (PSS + RS) (2,000 queries) & 91.57\% & 97.72\%   \\ \hline
	
\end{tabular}
\caption{Misclassification rates of the Google oracle on adversarial samples ($\varepsilon=0.3$) produced with DNN and LR substitutes after $\rho=\{3,6\}$ augmentation iterations.. Substitutes are trained without and with refinements from Section~\ref{sec:learning-approximators}: periodic step size (PSS) and reservoir sampling (RS).}
\label{tbl:google-misclassification}
\end{table}

\vspace*{-0.05in}

The model trained using Google's machine learning service is a little more
robust to adversarial samples than the one trained using Amazon's service, but
is still vulnerable to a large proportion of samples: $88.94\%$ of adversarial
samples produced with a perturbation $\varepsilon=0.3$ using a LR substitute
trained with $800$ queries to the oracle are misclassified. This confirms the 
above demonstration of the feasibility of black-box attacks against the classifier hosted by Amazon. Furthermore, if we use PSS and RS, the 
misclassification rate is $91.57\%$ for the DNN substitute and $97.72\%$ for the LR substitute, which again
demonstrates that combining PSS and RS increases misclassification compared to the original method for $\rho=3$, and reduces by a factor of $3$ the number of queries ($2,000$) compared to the original method for $\rho=6$.

\vspace*{-0.05in}

\textbf{A brief discussion of defenses -} In an effort to evaluate possible defenses against such attacks, we now add these adversarial samples to the MNIST training dataset and train a
new instance of the classifier oracle with the same procedure. The new oracle
has an accuracy of $91.25\%$ on the MNIST test set. Adversarial samples crafted
by training a new DNN substitute, even without PSS and RS, are still misclassified at a rate of $94.2\%$ after
$\rho=3$ iterations and $100\%$ after $\rho=6$. This defense is thus not
effective to protect the oracle from adversaries manipulating inputs. This is
most likely due to the fact that the Google Prediction API uses shallow
techniques to train its machine learning models, but we have no means to verify this. One could
also try to deploy other defense mechanisms like defensive
distillation~\cite{papernot2015distillation}. Unfortunately, as we do not have
any control on the training procedure used by Google Cloud, we cannot do so. To the
best of our knowledge, Google has not disclosed the machine learning technique
they use to train models served by their Google Cloud Prediction API service.
As such, we cannot make any further recommendations on how to better secure
models trained using this service. 


\section{Adversarial Sample Crafting}
\label{sec:adv-sample-crafting}

This section describes machine learning techniques used in this paper, along
with methods used to \emph{craft} adversarial samples against classifiers
learned using these techniques.  Building on
previous work~\cite{szegedy2013intriguing,goodfellow2014explaining,papernot2015limitations}
describing how adversaries can efficiently select perturbations leading deep
neural networks  to misclassify their inputs, we introduce new  crafting
algorithms for adversaries targeting Support Vector Machines (SVMs) and
Decision Trees (DTs).

\subsection{Deep Neural Networks}

Deep Neural Networks (DNNs) learn hierarchical representations of high
dimensional inputs used to solve ML tasks~\cite{Goodfellow-et-al-2016-Book},
including classification. Each representation is modeled by a layer of
neurons---elementary parameterized computing units---behaving like a
multi-dimensional function. The input of each layer $f_i$ is the output of the
previous layer $f_{i-1}$ multiplied by a set of weights, which are part of the
layer's parameter $\theta_i$. Thus, a DNN $f$ can be viewed as a composition of
parameterized functions $$f:\vec{x}\mapsto
f_n(\theta_n,...f_2(\theta_2,f_1(\theta_1,\vec{x}))...)$$ whose parameters
$\theta=\{\theta_i\}_i$ are learned during training. For instance, in the case
of classification, the network is given a large collection of known input-label
pairs $(\vec{x},y)$ and adjusts its parameters $\theta$ to reduce the label
prediction error $f(\vec{x})-y$ on these inputs. At test time, the model
extrapolates from its training data to make predictions $f(\vec{x})$ on unseen
inputs.

\vspace*{-0.05in}

To craft adversarial samples misclassified by DNNs, an adversary with knowledge
of the model $f$ and its parameters $\theta$ can use the \emph{fast gradient
sign method} introduced in~\cite{goodfellow2014explaining} or the
Jacobian-based iterative approach proposed in~\cite{papernot2015limitations}.
We only provide here a brief description of the fast gradient sign method,
which is the one we use in this work. To find an adversarial sample $\vec{x^*}$
approximatively solving the optimization problem stated in
Equation~\ref{eq:adv-sample-crafting-misclassification}, Goodfellow et
al.~\cite{goodfellow2014explaining} proposed to compute the following
perturbation:
\begin{equation}
\label{eq:fgsm}
\delta_{\vec{x}} = \varepsilon\sgn (\nabla_{\vec{x}} c(f, \vec{x}, y))
\end{equation}
where $f$ is the targeted DNN, $c$ its associated cost, and $y$ the correct
label of input $\vec{x}$. In other words, perturbations are evaluated as the
sign of the model's cost function gradient with respect to inputs. An
adversarial sample $\vec{x^*}=\vec{x}+\delta_{\vec{x}}$ is successfully crafted
when misclassified by model $f$---it satisfies $f(\vec{x^*})\neq
f(\vec{x})$---while its perturbation $\delta_{\vec{x}}$ remains
indistinguishable to humans. The \emph{input variation} $\varepsilon$ sets the
perturbation magnitude: higher input variations yield samples more likely to be
misclassified by the DNN model but introduce more perturbation, which can be
easier to detect.

\subsection{Multi-class Logistic Regression}

Multi-class logistic regression is the generalization of logistic regression to
classification problems with $N>2$ classes~\cite{murphy2012machine}. Logistic
regression seeks to find the hypothesis best matching the data among the class
of hypothesis that are a composition of a sigmoid function over the class of
linear functions. A multi-class logistic regression model $f$ can be written
as:
\begin{equation}
\label{eq:logistic-regression}
f:\vec{x}\mapsto \left[\frac{e^{\vec{w_j} \cdot \vec{x}}}{\sum_{l=1}^{N}e^{\vec{w_l} \cdot\vec{x}}} \right]_{j\in 1..N}
\end{equation}  
where $\theta=\{w_1, ..., w_N\}$ is the set of parameters learned during training, e.g., by gradient descent or Newton's method.

\vspace*{-0.05in}

Adversaries can also craft adversarial samples misclassified by multi-class
logistic regression models using the fast gradient sign
method~\cite{goodfellow2014explaining}. In the case of logistic regression, the
method finds the most damaging perturbation $\delta_{\vec{x}}$ (according to
the max norm) by evaluating Equation~\ref{eq:fgsm}, unlike the case of deep
neural networks where it found an approximation.

\subsection{Nearest Neighbors}

The k nearest neighbor (kNN) algorithm is a lazy-learning non-parametric
classifier~\cite{murphy2012machine}: it does not require a training phase.
Predictions are made on unseen inputs by considering the $k$ points in the
training sets that are closest according to some distance. The estimated class
of the input is the one most frequently observed among these $k$ points. When
$k$ is set to $1$, as is the case in this paper, the classifier is:
\begin{equation}
\label{eq:knn}
f:\vec{x}\mapsto Y\left[\arg\min_{\vec{z}\in X} \|\vec{z}-\vec{x}\|_2^2\right]
\end{equation}
which outputs one row of $Y$, the matrix of indicator vectors encoding labels for the training data $X$.

\vspace*{-0.05in}
 
Although the kNN algorithm is non-parametric, it is still vulnerable to
adversarial samples as pointed out
in~\cite{papernot2016practical,WardeFarley16}.  In this paper, we used the fast
gradient sign method to craft adversarial samples misclassified by nearest
neighbors. To be able to differentiate the models, we use a \emph{smoothed}
variant of the nearest neighbor classifiers, which replaces the argmin
operation in Equation~\ref{eq:knn} by a soft-min, as follows:
\begin{equation}
\label{eq:knn}
f:\vec{x}\mapsto  \frac{ \left[e^{-\|\vec{z}-\vec{x}\|_2^2}\right]_{\vec{z}\in X} }{\sum_{\vec{z}\in X} e^{-\|\vec{z}-\vec{x}\|_2^2}}  \cdot Y
\end{equation}

\subsection{Multi-class Support Vector Machines}

One possible implementation of a multiclass linear Support Vector Machine
classifier $f$ is the \emph{one-vs-the-rest} scheme.  For each class $k$ of the
machine learning task, a binary Support Vector Machine classifier $f_k$ is
trained with samples of  class $k$ labeled as positive and samples from other
classes labeled as negative~\cite{bishop2006pattern}. To classify a sample,
each binary linear SVM classifier $f_k$ makes a prediction and the overall
multiclass classifier $f$ outputs the class assigned the strongest confidence.
Each of these underlying linear SVMs is a model $f_k$  classifying unseen
samples $\vec{x}$ using the following: 
\begin{equation}
\label{eq:sub-svm-binary}
f_k:\vec{x} \mapsto sgn(\vec{w}[k]\cdot \vec{x} + b_k) 
\end{equation}

\begin{figure}[t]
	\centering
	\includegraphics[width=0.9\columnwidth]{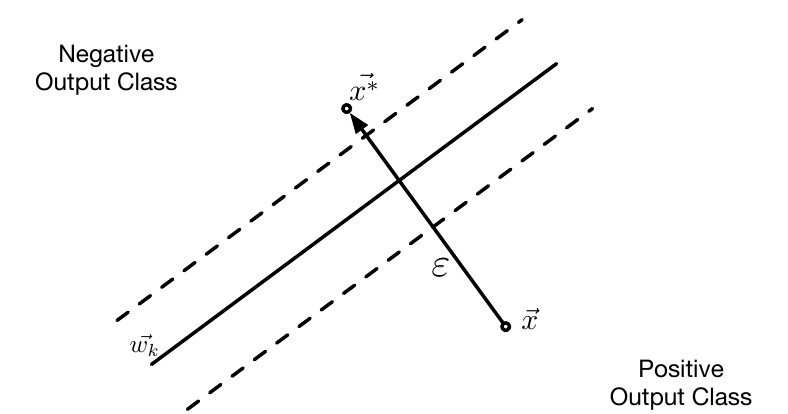}
	\caption{SVM Adversarial Samples: to move a sample $\vec{x}$ away from its legitimate class in a binary SVM classifier $f_k$, we perturb it by $\varepsilon$ along the direction orthogonal to $\vec{w[k]}$.}
	\label{fig:svm-adv-sample}
\end{figure}

We now introduce an algorithm to find adversarial samples misclassified by a multi-class linear SVM $f$. To the best of our knowledge, this method is more computationally efficient than previous~\cite{biggio2013evasion}: it does not require any optimization. To craft adversarial samples, we perturb a given input in a direction orthogonal to the decision boundary hyperplane. More precisely, we perturb legitimate samples correctly classified by model $f$ in the direction orthogonal to the weight vector $\vec{w}[k]$ corresponding to the binary SVM classifier $f_k$ that assigned the correct class $k$ output by the multiclass model $f$. The intuition, illustrated in Figure~\ref{fig:svm-adv-sample} with a binary SVM classifier, can be formalized as follows: for a sample $\vec{x}$ belonging to class $k$, an adversarial sample misclassified by the multiclass SVM model $f$ can be computed by evaluating:
\begin{equation}
\label{eq:svm-adv-sample}
\vec{x^*} = \vec{x} - \varepsilon \cdot \frac{\vec{w[k]}}{\|\vec{w_k}\|} 
\end{equation}
where $\|\cdot \|$ is the Frobenius norm, $\vec{w[k]}$ the weight vector of binary SVM $k$, and $\varepsilon$ the \emph{input variation} parameter. The input variation parameter controls the amount of distortion introduced as is the case in the fast gradient sign method. 
  
\subsection{Decision Trees}
Decision trees are defined by recursively partitioning the input
domain~\cite{murphy2012machine}. Partitioning is performed by selecting a
feature and a corresponding condition threshold that best minimize some cost
function over the training data. Each node is a if-else statement with a
threshold condition corresponding to one of the sample's features. A sample is
classified by traversing the decision tree from its root to one of its leaves
accordingly to conditions specified in intermediate tree nodes. The  leaf
reached indicates the class assigned.

\vspace*{-0.05in}

Adversaries can also craft adversarial inputs misclassified by decision trees.
To the best of our knowledge, this is the first adversarial sample crafting
algorithm proposed for decision trees. The intuition exploits the underlying
tree structure of the classifier model. To find an adversarial sample, given a
sample and a tree, we simply search for leaves with different classes in the
neighborhood of the leaf corresponding to the decision tree's original
prediction for the sample. We then find the path from the original leaf to the
adversarial leaf and modify the sample accordingly to the conditions on this
path so as to force the decision tree to misclassify the sample in the
adversarial class specified by the newly identified leaf. 

\begin{figure}[t]
	\centering
	\includegraphics[width=0.62\columnwidth]{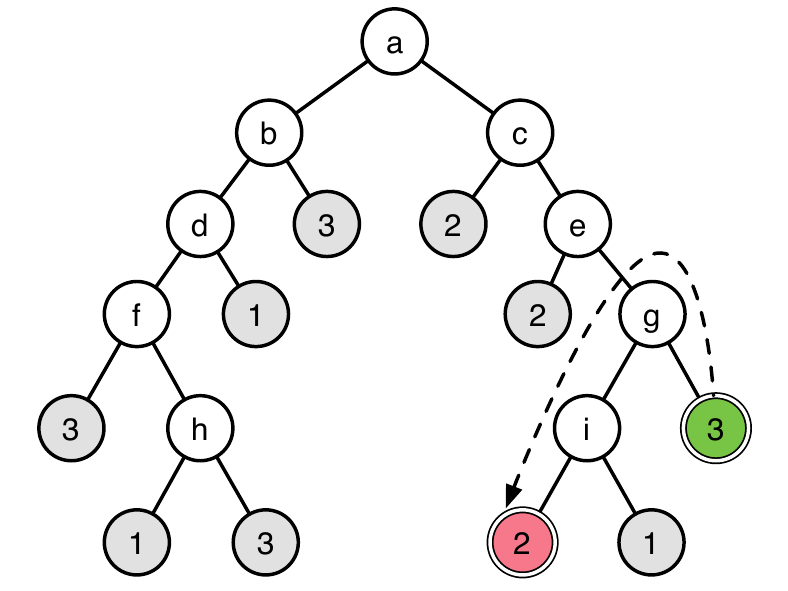}
	\caption{Decision Tree Adversarial Samples: leaves indicate output classes (here the problem has 3 output classes) whereas intermediate nodes with letters indicate binary conditions (if condition do else do). To misclassify the sample from class $3$  denoted by the green leaf, the adversary modifies it such that conditions $g$ and $i$ evaluate accordingly for the sample to be classified in class $2$ denoted by the red leaf.}
	\label{fig:decision-tree-adv-sample}
\end{figure}

This intuition, depicted in Figure~\ref{fig:decision-tree-adv-sample}, is
formalized by Algorithm~\ref{alg:decision-tree-adv-sample}. The algorithm takes
a decision tree $T$, a sample $\vec{x}$, the
$\small{\texttt{legitimate\_class}}$ for sample $\vec{x}$, and outputs an
adversarial sample $\vec{x^*}$ misclassified by decision tree $T$. The
algorithm does not explicitly minimize the amount of perturbation introduced to
craft  adversarial samples, but as shown in Section~\ref{sec:transferability},
we found in practice that perturbations found involve a minuscule proportion of
features.

\begin{algorithm}[t]
\caption{Crafting Decision Tree Adversarial Samples}
\label{alg:decision-tree-adv-sample}
\begin{algorithmic}[1]
	\Require $T$, $\vec{x}$, $\small{\texttt{legitimate\_class}}$
	\State $\vec{x^*} \leftarrow \vec{x}$
	\State $\small{\texttt{legit\_leaf}} \leftarrow$ find leaf in $T$ corresponding to $\vec{x}$
	\State $\small{\texttt{ancestor}} \leftarrow \small{\texttt{legitimate\_leaf}} $
	\State $\small{\texttt{components}} \leftarrow \small{\texttt{[]}}$
	\While{$\small{\texttt{predict}}(T,\vec{x^*}) == \small{\texttt{legitimate\_class}}$}
		\If{$\small{\texttt{ancestor}} == \small{\texttt{ancestor.parent.left}}$}
			\State $\small{\texttt{advers\_leaf}} \leftarrow$ find leaf under $\small{\texttt{ancestor.right}}$
		\Else{$\small{\texttt{ancestor}} == \small{\texttt{ancestor.parent.right}}$}
			\State $\small{\texttt{advers\_leaf}} \leftarrow$ find leaf under $\small{\texttt{ancestor.left}}$
		\EndIf
		\State $\small{\texttt{components}} \leftarrow $ nodes from $\small{\texttt{legit\_leaf}} $ to  $\small{\texttt{advers\_leaf}}$ 
		\State $\small{\texttt{ancestor}} \leftarrow \small{\texttt{ancestor.parent}}$
	\EndWhile
	\For{$i\in \small{\texttt{components}}$}
		\State perturb $\vec{x^*}[i]$ to change node's condition output
	\EndFor
	\State \Return $\vec{x^*}$
\end{algorithmic}
\end {algorithm}


\section{Discussion and Related Work}

Upon completion of their training on  collections of known input-label pairs
$(\vec{x},\vec{y})$, classifiers $f$ make label predictions $f(x)$ on unseen
inputs $\vec{x}$~\cite{murphy2012machine}. 
Models
extrapolate from knowledge extracted by processing input-label pairs during
training to make label predictions.  Several factors, including (1)
imperfections in the training algorithms, (2) the linearity of many underlying
components used to built machine learning models, and (3) the limited amount of
training points not always representative of the entire plausible input domain,
leave numerous machine learning models exposed to adversarial manipulations of
their inputs despite having  excellent performances on
legitimate---expected---inputs. 

Our work builds on a practical method for attacks against black-box deep
learning classifiers~\cite{papernot2016practical}. Learning substitute models
approximating the decision boundaries of targeted classifiers alleviates the
need of previous
attacks~\cite{szegedy2013intriguing,goodfellow2014explaining,papernot2015limitations}
for knowledge of the target architecture and parameters. We generalized this method and showed that it can target any machine
learning classifier. We also reduced its computational cost by (1)
introducing substitute models trained using logistic regression instead of deep
learning and (2) decreasing the number of queries made with reservoir sampling. Learning substitutes is an instance of knowledge transfer, a set of techniques to transfer the generalization knowledge learned by a model into another model~\cite{bucilua2006model, chen2015net2net}. 

\vspace*{-0.05in}
This paper demonstrates that adversaries can reliably target classifiers whose characteristics are unknown, deployed
remotely, e.g., by machine learning as a service platforms.  
The existence of such a threat vector calls for the design of defensive
mechanisms~\cite{mpc16}. Unfortunately, we found that defenses proposed in the
literature---such as training with adversarial
samples~\cite{goodfellow2014explaining}---were noneffective, or we were unable
to deploy them because of our lack of access to the machine learning model
targeted---for instance distillation~\cite{papernot2015distillation}. This
failure is most likely due to the shallowness of models like logistic
regression, which support the services offered by Amazon and Google, although
we are unable to confirm that statement in Google's case using available
documentation. 

\vspace*{-0.05in}

This work is part of a series of security evaluations of machine
learning algorithms~\cite{barreno2006can,biggio2014security}. Unlike us, previous work in this field assumed knowledge of the model
architecture and parameters~\cite{biggio2011support,huang2011adversarial}. Our threat model considered
adversaries interested in misclassification at test time, once the model has
been deployed. Other largely unexplored threat models exist. For
instance poisoning the training data used to learn models was only considered in the context of binary
SVMs whose training data is known~\cite{biggio2012poisoning} or anomaly detection systems whose underlying model is known~\cite{kloft2010online}.


\section{Conclusions}

Our work first exposed the strong phenomenon of adversarial sample 
transferability across the machine learning space. Not only do we 
find that adversarial samples are misclassified across models trained
using the same machine learning technique, but also across models
trained by different techniques. We then improved the accuracy and reduced the computational complexity of
an existing algorithm for learning  models substitutes of machine 
learning classifiers. We showed that DNNs and LR could both effectively
be used to learn a substitute model for many classifiers trained with
a deep neural network, logistic regression, support vector machine, 
decision tree, and nearest neighbors. In a final experiment, we 
demonstrated how all of these findings could be used to target online
classifiers trained and hosted by Amazon and Google, without any 
knowledge of the model design or parameters, but instead simply by making 
label queries for $800$ inputs. The attack successfully forces these
classifiers to misclassify $96.19\%$ and $88.94\%$ of their inputs.

\vspace*{-0.05in}

These findings call for some validation of inputs used by machine
learning algorithms. This remains an open problem. 
Future work should continue to improve the learning of substitutes to
maximize their accuracy and the transferability of adversarial samples
crafted to targeted models. Furthermore, poisoning 
attacks at training time remain largely to be investigated, leaving room 
for contributions to the field.




\section{Acknowledgments}

	Research was sponsored by the Army Research Laboratory and was accomplished
	under Cooperative Agreement Number W911NF-13-2-0045 (ARL Cyber Security
	CRA). The views and conclusions contained in this document are those of the
	authors and should not be interpreted as representing the official policies,
	either expressed or implied, of the Army Research Laboratory or the U.S.
	Government. The U.S. Government is authorized to reproduce and distribute
	reprints for Government purposes notwithstanding any copyright notation here
	on.

%

%
%




\end{document}